\documentclass[a4paper,fleqn]{cas-dc}
\usepackage[numbers,sort&compress]{natbib}
\usepackage{ulem}
\usepackage{enumitem}
\usepackage{graphicx} 
\usepackage{listings}
\usepackage{xcolor}
\usepackage{multirow}
\usepackage{algorithm}
\usepackage{algpseudocode}
\usepackage{makecell}
\usepackage[utf8]{inputenc}
\usepackage[most]{tcolorbox}
\usepackage{tikz}
\usetikzlibrary{trees}
\usepackage[strings]{underscore}
\usepackage{tikz}
\usepackage{fontawesome5} 
\usetikzlibrary{shapes,arrows.meta,positioning,calc,fit}
\usepackage{url}
\usepackage{booktabs}
\usepackage{hyperref}
\usepackage{float}
\usepackage{framed}
\usepackage{amssymb}
\usepackage{pifont}
\newcommand{\xmark}{\ding{55}}

\ExplSyntaxOn
\cs_gset:Npn \__first_footerline:
  { \group_begin: \small \sffamily \__short_authors: \group_end: }
\ExplSyntaxOff

\def\tsc#1{\csdef{#1}{\textsc{\lowercase{#1}}\xspace}}
\tsc{WGM}
\tsc{QE}
\tsc{EP}
\tsc{PMS}
\tsc{BEC}
\tsc{DE}

\tikzset{
  box/.style = {rectangle, draw=black, rounded corners, minimum width=28mm, minimum height=8mm, align=center, font=\sffamily},
  smallbox/.style = {rectangle, draw=black, rounded corners, minimum width=22mm, minimum height=7mm, align=center, font=\sffamily\footnotesize},
  arrow/.style = {-{Stealth[length=3mm]}, semithick},
  grouped/.style = {draw=black,dashed,rounded corners,inner sep=6pt},
  arrowicon/.style = {font=\small} 
}

\definecolor{addedB}{RGB}{255,165,0}

\def\categoryCount{11}
\def\newCategoryCount{4}

\def\buildCount{5}
\def\defectCount{90}

\def\onHoldCount{47}

\def\designCount{16}

\def\documentationCount{20}

\def\uncertaintyCount{69}

\def\lowInternalQualityCount{49}

\def\otherCount{225}
\def\skipTestCount{39}

\def\workaroundCount{66}

\def\codeCount{115}

\def\newTypeCount{133}

\def\limitedTestCount{26}

\def\totalSatdCount{615}

\def\repositoryUrl{https://github.com/SQMLab/TEST-SATD/blob/main}

\definecolor{commentColor}{rgb}{0.0, 0.35, 0.0}

\begin{document}

\title[mode=title]{A First Look at the Self-Admitted Technical Debt in Test Code: Taxonomy and Detection}
\shorttitle{A First Look at the Self-Admitted Technical Debt in Test Code: Taxonomy and Detection}

\shortauthors{Islam et al.}
\author[sqm]{Shahidul Islam}
\ead{islams32@myumanitoba.ca}

\author[sqm]{Md Nahidul Islam Opu}
\ead{opumni@myumanitoba.ca}

\author[mamba]{Shaowei Wang}
\ead{Shaowei.Wang@umanitoba.ca}

\author[sqm]{Shaiful Chowdhury}
\ead{shaiful.chowdhury@umanitoba.ca}

\affiliation[sqm]{%
  organization={SQM Research Lab, Computer Science, University of Manitoba},
  city={Winnipeg},
  state={MB},
  country={Canada}
}
\affiliation[mamba]{%
  organization={MAMBA Lab, Computer Science, University of Manitoba},
  city={Winnipeg},
  state={MB},
  country={Canada}
}


\begin{abstract}
Self-admitted technical debt (SATD) refers to comments in which developers explicitly acknowledge code issues, workarounds, or suboptimal solutions. SATD is known to significantly increase software maintenance effort. While extensive research has examined SATD in source code, its presence and impact in test code have received no focused attention, leaving a significant gap in our understanding of how SATD manifests in testing contexts.

This study investigates SATD in test code by manually analyzing 50,000 comments randomly sampled from 1.6 million comments across 1,000 open-source Java projects. From this sample, after manual analysis and filtering, we identified \totalSatdCount{} SATD comments and classified them into \categoryCount{} distinct categories, building a taxonomy of test code SATD. To investigate whether test code SATD can be detected automatically, we evaluated existing SATD detection tools, as well as both open-source and proprietary LLMs. Among the existing tools, MAT performed the best, albeit with moderate recall. To our surprise, both open-source and proprietary LLMs exhibited poor detection accuracy. These results indicate that neither existing approaches nor current LLMs can reliably detect SATD in test code.

Overall, this work provides the results of a large-scale analysis of SATD in test code, a nuanced understanding of its types, and the limitations of current SATD detection methods. Our findings lay the groundwork for future research on test code-specific SATD.
\end{abstract}






\maketitle
\section{Introduction}

Software maintenance keeps applications reliable and adaptable as they evolve. It involves continuous updates to fix bugs, improve performance, and meet user needs, often demanding more effort than initial development \cite{kafura_use_1987, borstler_role_2016}. As such, to reduce future maintenance burden, researchers have developed methods to identify and manage code components that are more maintenance prone~\cite{pascarella_performance_2020, chowdhury_method-level_2024, romano_using_2011, chowdhury_revisiting_2022, zhou_ability_2010,grund_codeshovel_2021, palomba_exploratory_2017}.
Previous research on software maintenance, however, has predominantly focused on production code (i.e., source code), with significantly less attention given to test code~\cite{winkler_investigating_2024, tahir_empirical_2016}. This oversight is significant, as the quality and maintainability of test code also play a crucial role in overall software quality and long-term maintainability~\cite{spadini_relation_2018, zaidman_studying_2011, vahabzadeh_empirical_2015}. 

Consider the case of Self-Admitted Technical Debt (SATD)~\cite{potdar_exploratory_2014, wehaibi_examining_2016, seaman_technical_2015, zazworka_investigating_2011, spinola_investigating_2013, guo_tracking_2011,xiao_quantifying_2024,cassee_self-admitted_2022}, where developers explicitly acknowledge the need for improvement or rework through comments in the source code. SATD has been widely recognized as a factor that negatively impacts software maintainability. For example, a recent study by Chowdhury \textit{et al.}~\cite{chowdhury_evidence_2025} found that source code methods containing SATD not only exhibit lower code quality, but are also more change-prone and bug-prone than their non-SATD counterparts. Not surprisingly, considerable research has focused on detecting~\cite{maldonado_detecting_2015, liu_satd_2018, huang_identifying_2018, li_automatic_2023, sheikhaei_empirical_2024, maldonado_using_2017,gama_towards_2024, rantala_prevalence_2020}, classifying~\cite{zhu_detecting_2021, maldonado_detecting_2015, obrien_artifact_2022, bhatia_empirical_2025}, and mitigating SATD~\cite{maldonado_empirical_2017, zampetti_was_2018, zampetti_automatically_2020, iammarino_empirical_2021, liu_exploratory_2021, mastropaolo_towards_2023}. Unfortunately, the subject samples in these studies have been either exclusively source code or overwhelmingly dominated by it.

Relying on findings about SATD in source code to understand SATD in test code may lead to inaccurate conclusions. This is because test code often exhibits characteristics that differ significantly from those of production code. For instance, test code tends to contain test-specific code smells~\cite{bavota_are_2015, tufano_empirical_2016}, a higher prevalence of code cloning~\cite{van_bladel_comparative_2023}, and a lower frequency of comments~\cite{counsell_comparing_2016}. Moreover, developers typically devote less attention to the quality of test code than to production code, which can result in a range of quality issues within test suites~\cite{counsell_comparing_2016, beller_developer_2019, beller_when_2015, zaidman_studying_2011, athanasiou_test_2014, spadini_relation_2018}. These differences highlight the need to study SATD in the test code independently. Motivated by this gap, our work focuses specifically on the detection and classification of SATD in test code.

 To study test code SATD, we collected over 1.6 million test code comments from 1,000 open-source Java projects. From this corpus, 50,000 comments were randomly sampled for manual labeling. We evaluated the performance of existing SATD detection tools on test code comments and investigated the effectiveness of Large Language Models (LLMs) for automated SATD detection. 
 In general, the contribution of this paper is founded on four research questions.

\textbf{RQ1: What are the different types of SATD in test code?}

After manually analyzing 50,000 comments and applying a filtering approach, we identified 615 SATD comments and classified them into \categoryCount{} distinct categories. While some of these categories—such as code debt and design debt—are consistent with those previously identified in source code SATD~\cite{maldonado_detecting_2015}, four categories are unique to test code. An example is the limited test category, which captures SATD specific to incomplete or low-value test validation.

\textbf{RQ2: Can existing SATD detection tools detect test code SATD?}

Test code comments exhibit patterns and intentions that differ from those found in source code. As a result, the performance of the existing SATD detection tools remains unknown. In this study, we evaluated seven existing approaches: pattern-based approach by Potdar et al.~\cite{potdar_exploratory_2014}, Natural Language Processing (NLP)-based approach by Maldonado et al.~\cite{maldonado_using_2017}, Text Mining (TM)-based approaches by Huang et al.~\cite{huang_identifying_2018} and Liu et al.~\cite{liu_satd_2018} (SATD Detector), Matches task Annotation
Tags (MAT)-based approach by Guo et al.~\cite{guo_how_2021}, a multi-task learning model based on BERT proposed by Gu et al.~\cite{gu_self-admitted_2024}, and DebtHunter by Sala et al.~\cite{sala_debthunter_2021}. We found that the MAT-based approach exhibits the best performance (0.75 F1-score), but with moderate recall.

\textbf{RQ3: Can open-source LLMs detect SATD in test code?}

Since SATD comments are written in natural language, LLMs may be better suited to detect them, particularly when comments lack recurring SATD keywords. Encouraged by Sheikhaei et al.~\cite{sheikhaei_empirical_2024}, we evaluated the Flan-T5 series (Small to XXL) under 0 to 4-shot settings using five different prompts. Results show that open-source LLMs struggle to reliably detect SATD in test code. While the best F1-scores reached 0.68 under MAT keywords suggested prompt, the recall was only 0.52.

\textbf{RQ4: Can proprietary LLMs detect SATD in test code?}

As proprietary LLMs can outperform open-source models in many scenarios~\cite{mayer_can_2024}, we evaluated two proprietary families—Gemini and GPT—under few-shot settings of 0, 2, and 4 shots. Interestingly, in contrast to the open-source LLMs, the tested proprietary LLMs exhibited very high recall but very low precision. 

In summary, we developed a taxonomy of test code SATD categories, revealing diverse maintenance issues. Additionally, we found that traditional tools, although they do not exhibit very good performance on test code SATD detection, their performance is surprisingly better than both open-source and proprietary LLMs. These findings highlight the need for test-aware SATD detection methods that can improve both precision and recall. 

To facilitate replication and extension, we share our dataset publicly.\footnote{\url{https://github.com/SQMLab/technical-debt}} This dataset can work as a benchmark for evaluating the accuracy of future detection and classification models and tools.  

The remainder of this paper is organized as follows. Section~\ref{sec:related_work} discusses related work on SATD. Section~\ref{sec:methodology} describes our data collection and manual SATD labeling process. Section~\ref{sec:approach_results} presents manual SATD classification, detection approaches, and the results for all research questions. Section~\ref{sec:discussion} discusses the implications of our findings, while Section~\ref{sec:threats} outlines threats to validity. Finally, Section~\ref{sec:conclusion} concludes the paper and highlights future directions.





\section{Related Work}\label{sec:related_work}
Software quality and maintainability have been extensively studied by the research community, covering a wide range of topics such as code quality metrics~\cite{mccabe_complexity_1976, chidamber_metrics_1994, mo_decoupling_2016}, change-proneness~\cite{chowdhury_revisiting_2022, khomh_exploratory_2012, romano_using_2011}, bug-proneness~\cite{chowdhury_method-level_2024, pascarella_performance_2020, giger_method-level_2012}, code readability~\cite{posnett_simpler_2011, buse_learning_2010}, and code understandability~\cite{scalabrino_automatically_2017}, among others. Due to its relevance to software quality and maintainability, a different stream of research has focused on SATD detection, classification, and its long-term maintenance impact. 

Potdar and Shihab~\cite{potdar_exploratory_2014} were the first to systematically study SATD in four large open-source projects, without differentiating between production and test code. They found SATD in 2.4–31.0\% of project files and investigated the reasons for its introduction and its removal frequency. Bavota and Russo~\cite{bavota_large-scale_2016} replicated the study of Potdar and Shihab, but by analyzing a large number of projects (159). They found that SATD comments tend to increase over time and persist for long periods. Chowdhury et al.~\cite{chowdhury_evidence_2025} conducted a large-scale method-level analysis of 774,051 methods across 49 open-source projects, revealing that methods with SATD are larger, more complex, less readable, and more prone to bugs and frequent changes. Similarly, Wehaibi et al.~\cite{wehaibi_examining_2016} analyzed five major open-source systems and found that technical debt negatively affects software evolvability, making future changes more difficult. Collectively, these studies underscore the negative impact of SATD on long-term software health, highlighting the importance of its classification and detection.

The detection of SATD has been explored extensively through both manual and automated approaches. Potdar and Shihab~\cite{potdar_exploratory_2014} identified 62 frequent keywords showing their potential in SATD detection. Various automated techniques have been proposed as well. Guo et al.~\cite{guo_how_2021} introduced the Matches task Annotation Tags (MAT) method, which achieves performance comparable to NLP-based approaches~\cite{maldonado_using_2017}. Huang et al.~\cite{huang_identifying_2018} applied text mining to extract features, performed feature selection, and trained a composite classifier for SATD detection. More recently, large language models (LLMs)~\cite{sheikhaei_empirical_2024, lambert_identification_2024,gu_self-admitted_2024} have been employed, outperforming traditional methods for detecting SATD in application code.

To better conceptualize technical debt, Alves et al.~\cite{alves_towards_2014} proposed an ontology synthesizing definitions and indicators scattered across the literature, categorizing technical debt into 13 types (e.g., architecture, build, code, test). Building on this foundation, Maldonado and Shihab~\cite{maldonado_detecting_2015} manually reviewed 33,093 heuristically filtered source code comments from five Java projects and identified 2,457 SATD comments, which they classified into five categories: design, requirement, defect, test, and documentation. Maldonado et al.~\cite{maldonado_using_2017} later extended this study to additional projects and applied natural language processing (NLP) techniques for classification. Their dataset has been used by subsequent studies on automated SATD detection and classification \cite{ sheikhaei_empirical_2024, guo_how_2021, gu_self-admitted_2024, arcelli_fontana_binary_2025, fucci_waiting_2021, njeru_hybrid_2025}.
Recent work has expanded SATD research into new domains beyond traditional software projects. OBrien et al.~\cite{obrien_artifact_2022} categorized SATD comments in machine learning code into 23 distinct types, while Bhatia et al.~\cite{bhatia_empirical_2025} extended Bavota et al.’s~\cite{bavota_large-scale_2016} taxonomy by adding new machine-learning-specific categories and subcategories to classify SATD in machine learning software. Liu et al.~\cite{liu_is_2020} examined seven popular open-source deep learning frameworks to assess SATD prevalence and categories. Pinna et al.~\cite{pinna_investigation_2023} found that SATD in blockchain projects is more prevalent than in non-blockchain systems. Azuma et al.~\cite{azuma_empirical_2022} manually classified SATD comments in Dockerfiles, identifying five main classes—including Docker-specific categories—to capture domain-specific debt. Similarly, Wilder et al.~\cite{wilder_exploratory_2023} analyzed 15,614 open-source Android applications and found that code debt is the most prevalent type.

Although a study by Counsell and Swift~\cite{counsell_timing_2022} examined SATD in test code in addition to source code, their focus was on understanding whether the time of day is related to writing SATD, rather than on identifying patterns of test code SATD or supporting their detection. During the course of our study, we became aware of a parallel work by Nakamura et al.~\cite{nakamura2025understanding}, which also investigated SATD in test code. However, our study differs substantially from theirs in several important aspects, as outlined below.

\begin{itemize}
     \item To identify test-code SATD, we manually examined 50,000 code comments. In contrast, the other study relied on an existing SATD detection tool, DebtHunter, which was not specifically designed or trained to detect SATD in test code. Our evaluation of DebtHunter on test-code SATD detection revealed significantly lower accuracy (RQ2). As such, using DebtHunter to construct ground-truth datasets, as done in ~\cite{nakamura2025understanding}, may have introduced inaccuracies into the identified SATD instances and affected the reliability of the resulting analysis.
     \item While constructing their taxonomy of test-code SATD, Nakamura et al. did not attempt to align their findings with existing SATD taxonomies. Such comparisons are important because they help determine how newly identified categories differ from or extend established ones, and this practice is commonly followed in software engineering research. For example, in their work, ``Why Attention Fails: A Taxonomy of Faults in Attention-Based Neural Networks'', Jahan et al.~\cite{jahan2025attention} introduced new categories only when the identified issues could not be adequately mapped to an existing deep-learning bug taxonomy. Similarly, in our study, we first attempted to map each SATD instance to the existing source-code SATD taxonomy and introduced new categories only when the comments could not be appropriately classified within the existing structure. Without such an analysis, it remains unclear whether a separate taxonomy for test-code SATD is necessary or whether existing SATD taxonomies are already sufficient to capture these instances.
 \item We argue that the problem of identifying whether a comment constitutes SATD should be addressed before attempting to classify it into specific SATD categories. If SATD comments cannot be reliably distinguished from non-SATD comments, then classifying their types becomes considerably less meaningful. However, the competing study focused on the multiclass classification task (i.e., predicting SATD categories) without first addressing the more fundamental binary classification task of distinguishing SATD from non-SATD comments. This concern is particularly significant because our results show that existing SATD detection tools, as well as both open-source and proprietary LLMs, struggle to accurately identify SATD in test code. These findings suggest that reliable test code SATD detection remains an open challenge and should be established before pursuing prediction of SATD classes.
 \item In addition to evaluating a wide range of open-source and proprietary LLMs, we also evaluated state-of-the-art SATD detection tools. This comparison is motivated by the practical consideration that LLM-based solutions can be computationally expensive, and therefore should ideally be used only if simpler approaches do not provide sufficient performance.

\end{itemize}

\begin{tcolorbox}
\textbf{\underline{Motivation}:}
While numerous studies have examined different aspects of test code maintenance \cite{spadini_relation_2018, lin_quality_2019, beller_developer_2019, beller_when_2015, zaidman_studying_2011, athanasiou_test_2014}, investigating the detection and classification of SATD in test code remains underexplored. Motivated by this gap, we conduct an empirical study on the test code SATD, exploring its types and detection.
\end{tcolorbox}






\begin{figure*}[!ht]
\centering
\begin{tikzpicture}[node distance=12mm and 14mm, every node/.style={font=\small}]

\node[box] (trending_project) {\faGithub\ 1K GitHub\\Trending Java Projects};
\node[box,right=1.5cm of trending_project] (comments) {\faComments\ 1.6 millions test\\code comments};
\node[box,right=2cm of comments] (analyzed_comments) {\faUserCheck\ 50K manually\\analyzed comments};
\node[box,right=1.8cm of analyzed_comments] (satd_comments) {\faComments\ 615 SATD\\comments};
\node[box, below=1.1cm of satd_comments] (classified_satd) {\faChartBar\ \categoryCount{} types of SATD};

\node[smallbox, below=22mm of analyzed_comments, xshift=-30mm] (rq2) {\faSearch\ Detect with\\existing tool};
\node[smallbox, below=22mm of analyzed_comments] (rq3) {\faBrain\ Detect with\\ open-source LLMs};
\node[smallbox, below=22mm of analyzed_comments, xshift=30mm] (rq4) {\faLock\ Detect with\\proprietary LLMs};

\draw[arrow] (trending_project) -- (comments) 
    node[midway, above, arrowicon]{\faTools} 
    node[midway, below, align=center]{extract\\comments};
\draw[arrow] (comments) -- (analyzed_comments) 
    node[midway, above, arrowicon]{\faRandom}
    node[midway, below, align=center]{randomly\\select and analyze};
\draw[arrow] (analyzed_comments) -- (satd_comments)
    node[midway, above, arrowicon]{\faFilter}
    node[midway, below, align=center]{filter};

\draw[arrow] (satd_comments) -- (classified_satd)
    node[midway, left, arrowicon]{\faUsers}
    node[midway, right, align=center]{RQ1\\manually\\classify};

\coordinate (branch) at ($(analyzed_comments.south)+(0,-1cm)$);

\draw[semithick] (analyzed_comments.south) -- (branch)
    node[midway, left, arrowicon]{\faRobot}
    node[midway, right, align=center]{automated detection};

\draw[arrow] (branch) -| (rq2.north)
    node[pos=0.3, below, align=center]{RQ2};
\draw[arrow] (branch) -- (rq3.north)
    node[midway, right]{RQ3};
\draw[arrow] (branch) -| (rq4.north)
    node[pos=0.3, below]{RQ4};

\end{tikzpicture}
\caption{Workflow for collecting, sampling, analyzing, classifying test code comments (RQ1), and evaluating automated SATD detection using existing tools (RQ2), open-source LLMs (RQ3), and proprietary LLMs (RQ4).}
\label{fig:methodology}
\end{figure*}

\section{Methodology}

\label{sec:methodology}
Figure~\ref{fig:methodology} presents a high-level overview of our overall methodology, outlining the key stages involved in data collection, comment extraction, manual labeling, and SATD detection using both traditional and LLM-based approaches. It also clarifies how each branch maps to our research questions: manual taxonomy construction in RQ1 and automated detection pipelines in RQ2--RQ4.

\subsection{Project Selection}
We aimed to collect comments from a diverse range of projects to avoid relying on a limited set of contributors, which could hinder the generalizability of our observations. Such generalizability is essential for developing a benchmark dataset, as selecting SATD from only a few projects would be strongly influenced by the commenting styles of a small number of individuals. This, in turn, would hinder the meaningful evaluation of current and future SATD detection tools. 
While selecting a large number of projects, however, we had to ensure that the collected comments were not predominantly drawn from low-quality projects (e.g., toy projects)~\cite{kalliamvakou_promises_2014}. Therefore, we used GitHub’s trending project list, ranked by the number of stars retrieved via the GitHub API. Although Borges et al.~\cite{borges_whats_2018} showed that star counts do not always correspond to sustained engagement or project quality, they also found that many practitioners continue to consider star counts before adopting a tool in their projects. Nevertheless, we mitigated the potential negative impact of star-based project selection by including another popularity indicator---we did not include projects with fewer than 100 forks. In addition, we randomly analyzed 100 projects and found that the selected repositories were well-maintained. 

\begin{figure*}[!ht]
\centering
\includegraphics[width=\textwidth]{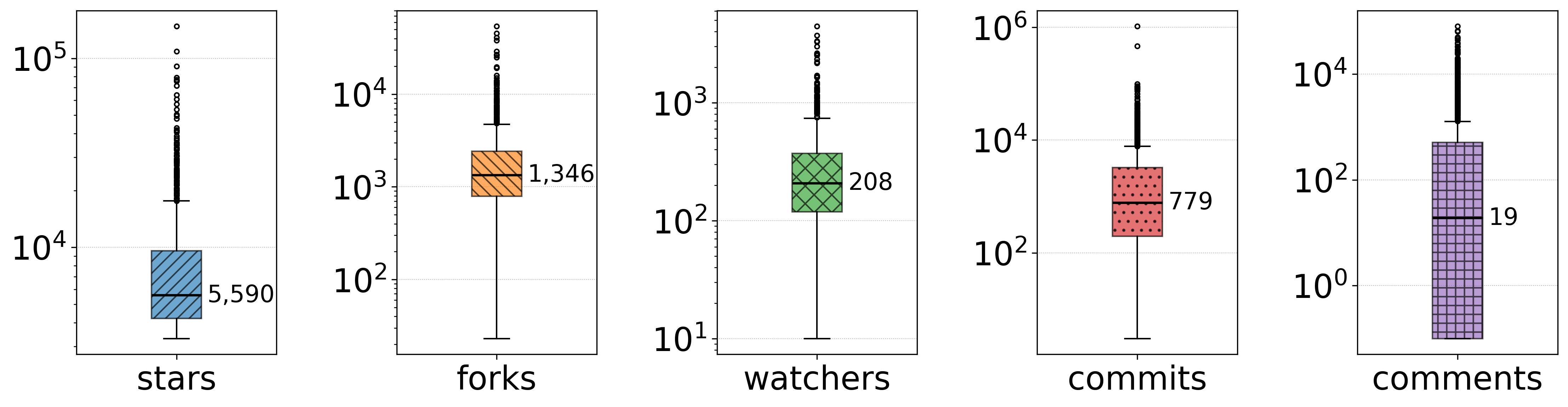}
\caption{Distribution of stars, forks, watchers, commits, and the number of extracted test code comments across the selected projects.}
\label{fig:repository_statistics}
\end{figure*}

At the end, we selected 1,000 Java-based repositories that were more likely to contain meaningful discussions and development practices relevant to our study. We focused on a single programming language only due to reliance on a parser for comment extraction—supporting multiple programming languages would require multiple parsers. Also, since our focus is on test code only, we needed techniques to exclude production code files (described in Section~\ref{sec:comment-parsing}), which would have been more challenging if applied across multiple programming languages. The selected projects cover a wide range of software types, including libraries, frameworks, applications, and middleware. Figure \ref{fig:repository_statistics} shows the distribution of stars, forks, watchers, commits, and extracted test code comment counts across these projects.

\subsection{Comment Parsing}
\label{sec:comment-parsing}
The source code of each project was first retrieved from GitHub with an automated script. Since this study focuses solely on test code, it is necessary to distinguish comments in production code from those in test code. To identify test code, we leveraged two heuristics. First, as is common in Java projects using Maven, Gradle, or Ant build systems, test code files and packages are typically placed under the src/test/java directory. Therefore, all Java files within this directory were treated as test code. Second, for files located outside the standard test directory, we checked for the presence of unit test annotations (e.g., @Test, @Before) from well-known unit testing frameworks such as JUnit, Mockito, and TestNG.

For code parsing, we used JavaParser\footnote{https://github.com/javaparser/javaparser last accessed: Oct 01, 2025} to extract all associated comments—including block, line, and Javadoc types—along with their start and end line numbers. To preserve the continuity of comments, we concatenated consecutive single-line comments. This is important because developers often use multiple single-line comments to document a single concept, and failing to merge them would result in fragmented sentences and a loss of contextual meaning. As a result, the raw number of extracted comments does not directly correspond to the number of semantically distinct comments. After merging consecutive single-line comments, we obtained a total of 1,653,658 test code comments, which were stored in a relational database to enable efficient downstream processing. Among the 1,000 analyzed projects, 45 contained a substantial number of comments, each exceeding 10,000. An additional 140 projects had between 1,000 and 10,000 comments, while 497 projects had fewer than 1,000 comments. The remaining 318 projects contained no test code comments at all.

\subsection{Dataset Creation}
Since manually reviewing more than 1.6 million test comments to identify SATD comments will be extremely time-consuming, we randomly selected 50,000 comments. Similar to the original SATD study~\cite{potdar_exploratory_2014}, the first author, a graduate student with 10+ years of industry experience as a lead software engineer, checked how many of these comments are SATD, making two classes of comments: SATD and non-SATD. This process required approximately 165 hours to complete. To assess labeling reliability, we conducted an independent double-check using a statistically grounded sample targeting a 95\% confidence level. The first author provided the second author with 619 randomly selected comments, obtained through stratified sampling across the two classes (237 SATD and 382 non-SATD). The second author, another graduate student with 2+ years of industry experience as a software engineer, independently labeled these 619 samples. The Cohen's kappa coefficient~\cite{cohen_coefficient_1960} was 0.95, implying an almost perfect~\cite{viera2005understanding} inter-rater agreement score and the reliability of our dataset labeling.   


Our manual review identified 1,024 SATD comments. However, among these 1,024 SATD comments, a striking 409 originated from a single project, OpenAPI Generator, accounting for approximately 38\% of that project’s comments. Upon closer inspection, we found that this repository contains hundreds of autogenerated Java client samples (e.g., OkHttp, Retrofit) with templated SATD comments following the pattern \textit{TODO}. Because this project disproportionately skewed the overall distribution of SATD across projects, including its comments would have hindered the generalizability of our benchmark dataset. For example, the taxonomy of SATD developed in this paper (RQ1) would be affected, and any evaluation of SATD detection tools would be biased. 
Consequently, we excluded all 1,063 comments from this project from our experimental dataset. We then removed 943 non-English comments. The refined dataset ultimately comprises 47,994 comments from 488 distinct projects, of which 615 (1.28\%) are classified as SATD, originating from 146 projects.



We also found that our filtered dataset of 47,994 comments contains 15,439  duplicate comments. Therefore, we created a second dataset applying deduplication. This allows us to better evaluate the model’s ability to make predictions on completely unseen comments. Thus, the original dataset (the duplicate set) preserves the natural distribution of comments as found in the projects, while the latter dataset allows us to evaluate detection effectiveness without the risk of any data leakage. Clearly, and unsurprisingly, our dataset is highly imbalanced, with SATD instances comprising only 1.28\% of the total data. As we show later, this severe imbalance makes accurate detection particularly challenging. 

Some of our evaluations in RQ2 involve retraining the existing tools. As such, we aimed to divide the dataset into 80\% training and 20\% testing. To prevent data leakage, we ensured that each project appeared exclusively in either the training or testing set, while also maintaining the proportional distribution of SATD and non-SATD comments across both sets---i.e., 20\% of all SATD should be in the test, and the rest in the train dataset. To do so,  we constructed the test set by randomly selecting one project at a time, measuring the distributions until approximately 20\% of the total SATD and 20\% of the total non-SATD comments were included. The remaining comments formed the training set. Because of project-wise selection, the split between training and testing was not always exactly 80/20, but it was very close. The details of these datasets are shown in Table~\ref{table:satd_detection_dataset}. In addition, for RQ2, we applied 5-fold cross-validation with stratified group-based splitting at the project level. This approach prevents data leakage~\cite{chowdhury_method-level_2024, pascarella_performance_2020} by ensuring that all samples from the same project remain within the same fold; thus, a project included in the training set cannot appear in the test set. Unless otherwise specified, we used the same test set for all settings to ensure a fair comparison among the evaluated approaches.

\begin{table*}[!ht]
\centering
\caption{Training and testing datasets. Percentages in parentheses indicate the proportion within the corresponding datasets. Please note that, to enable fair comparisons across all tools and approaches, all accuracy metrics in all RQs were reported using only the test dataset.}
\label{table:satd_detection_dataset}
\begin{tabular}{llrrrrr}
\toprule
\textbf{Distribution} & \textbf{Dataset}  & \textbf{Comments (\%)}& \makecell{\textbf{non-SATD} \\ \textbf{Comments (\%)}} &  \makecell{\textbf{SATD} \\ \textbf{Comments (\%)}} & \makecell{\textbf{Overall} \\ \textbf{SATD \%}} & \textbf{Projects (\%)}\\
\midrule

Original & Train & 38,380 (80.0) & 37,902 ( 80.0) & 478 (77.7) & 1.2 & 382 (78.3) \\
Original & Test & 9,614 (20.0) & 9,477 ( 20.0) & 137 (22.3) & 1.4 & 108 (22.1) \\
Deduplicate & Train & 26,024 (79.9) & 25,621 ( 80.0) & 403 (78.1) & 1.5 & 368 (79.0) \\
Deduplicate & Test & 6,531 (20.1) & 6,418 ( 20.0) & 113 (21.9) & 1.7 & 98 (21.0) \\

\bottomrule
\end{tabular}
\end{table*}

\subsection{Evaluation Metrics}

To evaluate SATD detection for RQ2–RQ4, we used Precision, Recall, and F1-score following prior SATD research~\cite{sheikhaei_understanding_2025, maldonado_using_2017, ren_neural_2019, guo_how_2021, gu_self-admitted_2024}. Additionally, because SATD comments comprise only 1.28\% of our dataset, we also report Matthews Correlation Coefficient (MCC), which is more informative for highly imbalanced classification settings. The metrics are calculated as follows.

\[
\text{Precision} = \frac{TP}{TP + FP}
\qquad \text{and} \qquad
\text{Recall} = \frac{TP}{TP + FN}
\]

\[
\text{F1-score} = \frac{2 \times \text{Precision} \times \text{Recall}}{\text{Precision} + \text{Recall}}
\]

\[
\text{MCC} = \frac{(TP \times TN) - (FP \times FN)}{\sqrt{(TP + FP)(TP + FN)(TN + FP)(TN + FN)}}
\]

Here, \( TP \) denotes true positives, \( TN \) true negatives, \( FP \) false positives, and \( FN \) false negatives.

For RQ3 and RQ4, we assessed whether performance differences among LLM configurations were statistically significant. We calculated repository-level F1-score and MCC values and conducted pairwise two-sided Wilcoxon signed-rank tests. We applied Holm~\cite{holm1979simple} correction to the resulting $p$-values to control for multiple comparisons, using a significance level of 0.05. We also calculated Cliff's delta ($\delta$) to quantify the direction and magnitude of each difference. Cliff's $\delta$ ranges from $-1$ to $1$ and, following Hess and Kromrey~\cite{hess2004robust}, its magnitude is considered negligible when $|\delta| < 0.147$, small when $0.147 \leq |\delta| < 0.33$, medium when $0.33 \leq |\delta| < 0.474$, and large when $|\delta| \geq 0.474$. Wilcoxon tests and Cliff's delta are widely adopted for non-parametric comparisons in empirical software engineering research~\cite{he2015empirical,chen2020savior,pecorelli2020testing,chowdhury2019greenscaler}.

\section{Approach and Results}\label{sec:approach_results}
In this section, we discuss the approach and results of each RQ. 

\subsection{RQ1: What are the different types of SATD in test code?}
\textbf{Motivation.} A categorization of SATD will provide a high-level understanding of the types of challenges developers encounter when writing and maintaining tests. This can help reveal which issues are most prevalent or critical in test code, and thus could be prioritized for maintenance and quality improvement efforts. We hypothesize that these categories may differ from those observed in source code, reflecting the unique role and structure of tests within the software development process. 

\textbf{Approach.} Manually categorizing SATD comments at a granular level is challenging for several reasons.
(i) Prior classifications sometimes distribute the same phenomenon across multiple categories. For instance, Alves et al.~\cite{alves_towards_2014} classify duplicated code as both code debt and design debt, creating ambiguity during assignment.
(ii) Certain comments can encompass more than one category. For example, the comment, \texttt{// TODO: tweak the tests when FLINK-13604 is fixed}, simultaneously reflects a dependent defect (awaiting a bug fix) and a requirement (a need to revise the tests). In other cases, interpretation may be subjective, such as with \texttt{// An unlikely exception — if this occurs, we can't proceed with the test}, which could reasonably be considered either a defect or a superficial test.
(iii) Many SATD comments are terse or lack context (e.g., a standalone \texttt{TODO}), making it difficult to infer the underlying issue without examining surrounding code.
(iv). Some comments describe scenarios or concerns not captured by current taxonomies, necessitating extensions to the existing classification scheme. For example, consider the comment \texttt{// Probably incorrect - comparing Object[] arrays with Arrays.equals}. This comment cannot be confidently classified as defect debt, nor does it closely align with other existing categories, since the developer expresses uncertainty about whether the issue is truly a defect. Given these challenges—including subjectivity, overlapping categories, and ambiguous content—we approached the categorization of the \totalSatdCount{} SATD comments in a structured manner by following established guidelines~\cite{alves_towards_2014,bavota_large-scale_2016} and joint classification sessions to reach consensus on ambiguous cases. Our goal was to preserve diverse categories rather than merge multiple types into broader labels, ensuring subtle distinctions in SATD were captured. 

Initially, the first, second, and last author jointly conducted three online meetings to examine each comment one by one, assessing the nature of the technical debt they represented. This process followed the guidelines proposed by Alves et al.~\cite{alves_towards_2014}, and Bavota and Russo~\cite{bavota_large-scale_2016}, assigning each comment to a specific category based on its characteristics. For each comment, we first attempted to assign it to one of the existing categories, as found from ~\cite{alves_towards_2014,bavota_large-scale_2016}. If a comment did not fit any existing category, we created a new category to reflect its unique characteristics. The first 200 comments were classified collaboratively by the three authors during those three meetings. However, by the time the first 142 comments had been reviewed, \categoryCount{} distinct categories had emerged, and no new category was found after reviewing the remaining 58 samples, possibly indicating that a saturation point was reached. 
The remaining \the\numexpr \totalSatdCount - 200 \relax{} comments were then categorized by the first author independently. In addition to 10+ years of industry experience, this author was also successful in manual labeling in recent times~\cite{islam2025historyfinder,opu2025understanding}.   
\begin{figure*}[!ht]
\resizebox{\textwidth}{!}{%
\centering
\begin{tikzpicture}[
  grow=down,
  level distance=14mm,
  edge from parent fork down,
  every node/.style = {draw,  fill=gray!20, align=center, 
  font=\small
  }, 
  edge from parent/.style = {draw, -latex, edge from parent fork down},
  level 1/.style={sibling distance=2.1cm}, 
  level 2/.style={sibling distance=16.5mm, level distance=18mm}
  ]
\node {SATD in test code\\(\totalSatdCount)}
  child[xshift=-1pt] {node {Code debt\\(\codeCount)}
    child {node[xshift=0pt] {Workaround\\(\workaroundCount)}}
    child {node[xshift=30pt] {Low internal quality\\(\lowInternalQualityCount)}}
  }
  child {node[xshift=-5pt] {Defect debt\\(\defectCount)}}
  child {node[xshift=0pt] {Documentation\\debt (\documentationCount)}}
  child {node[xshift=5pt] {Design debt\\(\designCount)}}
  child {node (build-debt) [xshift=4pt] {Build debt\\(\buildCount)}}
  child {node (test-debt) [fill=black!80, text=white, xshift=-5pt, inner xsep=6pt] {Test\\(\newTypeCount)}
    child {node (uncertainty-test) [fill=black!80, text=white, xshift=-65pt] {Uncertainty\\(\uncertaintyCount)}}
    child {node[fill=black!80, text=white, xshift=-55pt] {On-hold\\(\onHoldCount)}}
    child {node[fill=black!80, text=white, xshift=-50pt] {Skip test\\(\skipTestCount)}}
    child {node (limited-test) [fill=black!80, text=white, xshift=-37pt] {Limited test\\(\limitedTestCount)}}
  }
  child {node (other-debt) [xshift=-15pt] {Other debt\\(\otherCount)}};
\coordinate (boundary-top) at ($(build-debt.east)!0.5!(test-debt.west)$);
\coordinate (boundary-right) at ([xshift=7pt]other-debt.east);
\coordinate (boundary-left) at ([xshift=-7pt]uncertainty-test.west);
\coordinate (boundary-bend) at ($(test-debt.south)!0.25!(uncertainty-test.north)$);
\coordinate (boundary-bottom) at ([yshift=-7pt]limited-test.south);
\draw[red, densely dotted, line width=1pt]
  ([yshift=12pt]boundary-top |- build-debt.north)
  -- ([yshift=12pt]boundary-right |- other-debt.north)
  -- (boundary-right |- boundary-bottom)
  -- (boundary-left |- boundary-bottom)
  -- (boundary-left |- boundary-bend)
  -- (boundary-top |- boundary-bend)
  -- cycle;
\end{tikzpicture}
}
\caption{Manual categorization of the \totalSatdCount{} self-admitted technical debt instances in test code. The categories shown within the dotted red boundary contain SATDs that did not align with any of the existing categories found in the source-code SATD taxonomies we compared against. Because some SATD comments were assigned to more than one type, the count shown in a parent node is not necessarily the sum of the counts shown in its descendant nodes.}
\label{fig:satd-hierarchy}
\end{figure*}

\textbf{Results.} Figure~\ref{fig:satd-hierarchy} presents the complete hierarchy of these categories and their corresponding sub-categories found in the test code.  Four new groups of technical debt were identified, each with unique characteristics, and shown under \textit{Test}.  
The number of instances identified in each category is also shown within the respective rectangles. Each example presented in this section is hyperlinked to its original source file, with all corresponding links also provided in our replication package.\footnote{\url{https://github.com/SQMLab/technical-debt/blob/main/data/satd_comment_example.csv}}

\textbf{Code debt (\codeCount{}):} \textit{``Problems found in the source code which can negatively affect the legibility of the code, making it more difficult to be maintained''} \cite{alves_towards_2014}. Two sub-categories of this type of technical debt were identified. The first sub-category, \textit{workaround}, describes intentionally written suboptimal temporary fixes that are justified by the need to meet specific functional or non-functional requirements, often representing a trade-off between code quality and project goals~\cite{bavota_large-scale_2016}. Representative examples of this sub-category are provided below:
\begin{itemize}[itemsep=-2pt]
    \item \href{https://github.com/elastic/logstash/blob/5573b5ad777eef7a4204f2376bd601090d7866bc/logstash-core/src/test/java/org/logstash/instrument/metrics/gauge/RubyHashGaugeTest.java\#L49}{\textcolor{commentColor}{//hacky workaround using the toString method to avoid mocking the Ruby runtime}}
    \item \href{https://github.com/geoserver/geoserver/blob/8871198a41d255de8c964263f0a6f15d44d15d62/src/web/gwc/src/test/java/org/geoserver/gwc/web/layer/LayerCacheOptionsTabPanelTest.java\#L211}{\textcolor{commentColor}{// This is an utter hack but is the only way I could}}
    \item \href{https://github.com/opensearch-project/OpenSearch/blob/c82fffe8e333a8fff600dc22fa2e1aed7276d180/test/framework/src/main/java/org/opensearch/test/OpenSearchIntegTestCase.java\#L1278}{\textcolor{commentColor}{// this is just a temporary thing but it's easier to change if it is encapsulated.}}
    \item \href{https://github.com/apache/tomcat/blob/3d58c3ac70beb21588b120d21ff5c075801ca31e/test/org/apache/tomcat/websocket/TestWsSessionSuspendResume.java\#L198}{\textcolor{commentColor}{// Yes, a static variable is a hack.}}
    \item \href{https://github.com/eugenp/tutorials/blob/28ee6f8b9d8a6ee95ec0cef5d65a9891b6e1ed55/testing-modules/junit-4/src/test/java/com/baeldung/assume/ConditionallyIgnoreTestsUnitTest.java\#L59}{\textcolor{commentColor}{// This should use System.getProperty("os.name") in a real test.}}
\end{itemize}

The second sub-category, \textit{low internal quality}, refers to deficiencies in the source code, such as poor readability, misuse of programming constructs, or unnecessary complexity~\cite{bavota_large-scale_2016}. Instance of \textit{low internal quality} are:

\begin{itemize}[itemsep=-2pt]
    \item \href{https://github.com/google/closure-compiler/blob/c86e4dd27ef48f9aa49261b1391274023ce7b658/test/com/google/javascript/jscomp/TypedScopeCreatorTest.java\#L3689}{\textcolor{commentColor}{// TODO(sdh): Print a better name (https://github.com/\allowbreak google/closure-compiler/issues/2982)}}
    \item \href{https://github.com/elastic/elasticsearch/blob/3d457ef84ebf5d3ee0fd09339fbc3373ed5beb3c/server/src/test/java/org/elasticsearch/lucene/spatial/CartesianShapeDocValuesQueryTests.java\#L101}{\textcolor{commentColor}{// RandomIndexWriter is too slow here:}}
    \item \href{https://github.com/bazelbuild/bazel/blob/d7c71fa286b84a5dce8c60f1e8fb48295703fc61/src/test/java/com/google/devtools/build/lib/concurrent/PaddedAddressesTest.java\#L29}{\textcolor{commentColor}{// TODO: b/359688989 - clean this up}}
    \item \href{https://github.com/vespa-engine/vespa/blob/6782a126f815ab896e6c379caaf9e9e909aa5d09/vespaclient-container-plugin/src/test/java/com/yahoo/vespaxmlparser/MockReader.java\#L37}{\textcolor{commentColor}{//Ooooooo this is so ugly}}
\end{itemize}


\textbf{Defect debt (\defectCount{}):} \textit{``Defect debt consists of known defects, usually identified by testing activities or by the user and reported on bug track systems, that the CCB agrees should be fixed, but due to competing priorities, and limited resources have to be deferred to a later time''} \cite{alves_towards_2014}. Known, unfixed or partially fixed defects were identified, totaling \defectCount{} instances. Some examples include:
\begin{itemize}[itemsep=-2pt]
    \item \href{https://github.com/robolectric/robolectric/blob/d86ffdced16b5835a3135b156d651f1ae4b25819/integration_tests/nativegraphics/src/test/java/org/robolectric/shadows/ShadowNativeInterpolatorTest.java\#L63}{\textcolor{commentColor}{// known bug: time argument is unsigned 32bit in graphics library}}
    \item \href{https://github.com/google/error-prone/blob/4ae849ea7cea2cfb8b07b4842e0ca19dd91a2a1f/core/src/test/java/com/google/errorprone/bugpatterns/inlineme/InlinerTest.java\#L1266}{\textcolor{commentColor}{// TODO(b/189535612): this is a bug!}}
    \item \href{https://github.com/spring-cloud/spring-cloud-netflix/blob/0d4ddc660fdd79d898815e5a0f8beb66894fc4e9/spring-cloud-netflix-eureka-server/src/test/java/org/springframework/cloud/netflix/eureka/server/EurekaInstanceMonitorTests.java\#L44}{\textcolor{commentColor}{//FIXME: fix flaky test}}
    \item \href{https://github.com/netty/netty/blob/f3311e5db4c9d9559fac96bb2f48295d63f58b67/codec/src/test/java/io/netty/handler/codec/DefaultHeadersTest.java\#L264}{\textcolor{commentColor}{// This test doesn't work on midnight, January 1, 1970 UTC}}
\end{itemize}

\textbf{Uncertainty (\uncertaintyCount{}):} Stems from uncertainty or a lack of clarity regarding how to implement, test, or resolve an issue. This type of debt is often marked by open questions, speculative reasoning, or lack of understanding of the expected behavior. Illustrative examples are as follows:
\begin{itemize}[itemsep=-2pt]
    \item \href{https://github.com/sofastack/sofa-rpc/blob/be93251f7ac754178811dd04d826ebe8f0967a1a/test/test-integration/src/test/java/com/alipay/sofa/rpc/test/generic/bean/ComplexBean.java\#L84}{\textcolor{commentColor}{// Probably incorrect - comparing Object[] arrays with Arrays.equals}}
    \item \href{https://github.com/apache/beam/blob/2b18c28132630689639a308dcc30da1441fd65ae/sdks/java/extensions/sql/src/test/java/org/apache/beam/sdk/extensions/sql/meta/provider/kafka/KafkaTableProviderIT.java\#L303}{\textcolor{commentColor}{// I don't think doing this will be safe in parallel}}
    \item \href{https://github.com/spring-cloud/spring-cloud-gateway/blob/73185358b23df378e7f388a6fcd1e39b2f12007f/spring-cloud-gateway-server/src/test/java/org/springframework/cloud/gateway/config/GatewayRedisAutoConfigurationTests.java\#L41}{\textcolor{commentColor}{// TODO: figure out why I need these}}
    \item \href{https://github.com/pentaho/pentaho-kettle/blob/db5f39cc7d9f6223ef20e6da5102d5910e91af27/integration/src/it/java/org/pentaho/di/trans/steps/formula/Formula2IT.java\#L162}{\textcolor{commentColor}{// TODO: Is it correct?}}
    \item \href{https://github.com/OpenRefine/OpenRefine/blob/02c80d0e49bb11c0b21d62d47a5ca13cb11048a9/modules/grel/src/test/java/com/google/refine/expr/functions/arrays/JoinTests.java\#L42}{\textcolor{commentColor}{// TODO: This is current behavior, but is it what we want?}}
\end{itemize}

\textbf{On-hold (\onHoldCount{}):} Refers to cases where the test code relies on external components, such as libraries, APIs, or services, that hinder test execution due to incomplete support or limited functionality. Although in a different context, these types of dependency-induced SATD were also observed by Maipradit et al.~\cite{maipradit_wait_2020}. A selection of examples is given below:
\begin{itemize}[itemsep=-2pt]
    \item \href{https://github.com/oracle/graal/blob/009a6f93e563afb1d110b42dc5951b3d6b4156f8/truffle/src/com.oracle.truffle.api.test/src/com/oracle/truffle/api/test/polyglot/FileSystemsTest.java\#L2175}{\textcolor{commentColor}{/**
 * Setting file last access time does not work on JDK-20 on macOS with the hfs file system,
 * issue JDK-8298187.
 */}}
    \item \href{https://github.com/querydsl/querydsl/blob/dd905aebf987599044e20b5822e7f035cee6a868/querydsl-jdo/src/test/java/com/querydsl/jdo/JDOQueryStandardTest.java\#L120}{\textcolor{commentColor}{// NO list support in JDOQL
    \newline //        testData.listTests(store.products, otherStore.products, p);}}
    \item \href{https://github.com/elastic/elasticsearch/blob/3d457ef84ebf5d3ee0fd09339fbc3373ed5beb3c/x-pack/qa/oidc-op-tests/src/javaRestTest/java/org/elasticsearch/xpack/security/authc/oidc/C2IdOpTestCase.java\#L244}{\textcolor{commentColor}{// C2ID doesn't have a non JS login page :/, so use their API directly
    // see https://connect2id.com/products/server/\allowbreak docs/guides/login-page}}
    \item \href{https://github.com/openzipkin/zipkin/blob/e151986db4d1de1aea53618f18e4ee3acaa9ed3a/zipkin-collector/rabbitmq/src/test/java/zipkin2/collector/rabbitmq/RabbitMQExtension.java\#L74}{\textcolor{commentColor}{// mostly waiting for https://github.com/testcontainers/\allowbreak testcontainers-java/issues/3537}}
    
\end{itemize}

\textbf{Skip test (\skipTestCount{}):} Arises when certain tests are skipped or explicitly disabled, often due to environmental limitations or unresolved issues. A few examples are:
\begin{itemize}[itemsep=-2pt]
    \item \textcolor{commentColor}{/**
    * TODO: enable when window is supported.
    */}
    \item \href{https://github.com/apache/beam/blob/2b18c28132630689639a308dcc30da1441fd65ae/sdks/java/extensions/avro/src/test/java/org/apache/beam/sdk/extensions/avro/io/SerializableAvroCodecFactoryTest.java\#L107}{\textcolor{commentColor}{// Skip, zstd only supported for Avro 1.9+}}
    \item \href{https://github.com/apache/hadoop/blob/8424c155f11049e2d5a92b7f8d63d4c67d2eaaa6/hadoop-tools/hadoop-azure/src/test/java/org/apache/hadoop/fs/azure/TestNativeAzureFileSystemMocked.java\#L36}{\textcolor{commentColor}{// Ignore the following tests because taking a lease requires a real
    // (not mock) file system store. These tests don't work on the mock.}}
    \item \href{https://github.com/java-native-access/jna/blob/02bf7cc863eccc4067e9eb981f96b3cd8aabc00d/contrib/platform/test/com/sun/jna/platform/win32/Kernel32Test.java\#L1804}{\textcolor{commentColor}{// Testcase is disabled, as kernel32 ordinal values are not stable.
    // a library with a stable function <-> ordinal value is needed.}}
    \item \href{https://github.com/google/j2objc/blob/3664859a54dbe07905b2789cf06318c5b390e92f/jre_emul/android/platform/external/icu/android_icu4j/src/main/tests/android/icu/dev/test/format/NumberFormatTest.java\#L1942}{\textcolor{commentColor}{//  CLDR No Longer uses complex currency symbols.
    //  Skipping this test.}}
  
\end{itemize}

\textbf{Limited test (\limitedTestCount{}):} Refers to cases where tests provide incomplete or low-value validation, limiting their ability to establish confidence in the behavior under test. Limited tests appeared in two forms. In the first form, only a small, representative portion of the input space is tested, usually for the sake of time efficiency or resource constraints. Examples of this form are:
\begin{itemize}[itemsep=-2pt]
    \item \href{https://github.com/elastic/elasticsearch/blob/3d457ef84ebf5d3ee0fd09339fbc3373ed5beb3c/x-pack/plugin/rollup/src/test/java/org/elasticsearch/xpack/rollup/RollupJobIdentifierUtilTests.java\#L769}{\textcolor{commentColor}{// This only tests for calendar ordering, ignoring the other criteria}}
    \item \href{https://github.com/trinodb/trino/blob/b89da7e651fea169443b1aaedb0c406744fe2cff/plugin/trino-base-jdbc/src/test/java/io/trino/plugin/jdbc/BaseJdbcTableStatisticsTest.java\#L293}{\textcolor{commentColor}{// Not testing average length and min/max, as this would make the test less reusable and is not that important to test.}}
    \item \href{https://github.com/trinodb/trino/blob/b89da7e651fea169443b1aaedb0c406744fe2cff/core/trino-spi/src/test/java/io/trino/spi/type/TestVarchars.java\#L81}{\textcolor{commentColor}{// we do not check if the offset is in the middle of a code point}}
    \item \href{https://github.com/logisim-evolution/logisim-evolution/blob/4be1d129854cb333226464f994eac6da6e268dc8/src/test/java/com/cburch/logisim/LineBufferTest.java\#L455}{\textcolor{commentColor}{// FIXME: this test do not cover breaking remark into multiple lines}}
    \item \href{https://github.com/apache/doris/blob/3c1d7d27269cab0022bc5b293da45dbb6013dff0/fe/hive-udf/src/test/java/org/apache/doris/HllUDFTest.java\#L44}{\textcolor{commentColor}{// Theoretically, the HLL algorithm can count very large cardinality with little relative error rate,
    // which is less than 2\% in Doris. We cost about 5 minutes to test with one billion input numbers here
    // successfully, but in order to shorten the test time, we chose one million numbers.}}
\end{itemize}

In the second form, tests exercise code without meaningfully validating its functional correctness. In other words, this is testing for the sake of testing. Examples of this form include:


\begin{itemize}[itemsep=-2pt]
     \item \href{https://github.com/NationalSecurityAgency/ghidra/blob/db1b6860c9bbe8b659704cb3909c0bf54359fb7d/Ghidra/Framework/SoftwareModeling/src/test/java/ghidra/program/database/data/StructureDBTest.java\#L2087}{\textcolor{commentColor}{// TODO: these tests are too simple since they only replace undefined components}}
    \item \href{https://github.com/vespa-engine/vespa/blob/6782a126f815ab896e6c379caaf9e9e909aa5d09/messagebus/src/test/java/com/yahoo/messagebus/DynamicThrottlePolicyTest.java\#L98}{\textcolor{commentColor}{/**
 * Sort of a dummy test, as the conditions are perfect. In a more realistic scenario, below, the algorithm needs luck to climb this high.
 */}}
 \item \href{https://github.com/FasterXML/jackson-databind/blob/cfef132dbacbc6ed612051f81e88fb37c5632154/src/test/java/com/fasterxml/jackson/databind/introspect/JacksonAnnotationIntrospectorTest.java\#L146}{\textcolor{commentColor}{// kinda sorta wrong, but for testing's sake...}}
 \item \href{https://github.com/spring-projects/spring-framework/blob/6ded25ba28b11a1237c1a9053ddc1c91bfd31d71/spring-beans/src/test/java/org/springframework/beans/factory/support/ManagedListTests.java\#L77}{\textcolor{commentColor}{// doesn't make much sense in the context of a list...}}
\end{itemize}

\textbf{Documentation debt (\documentationCount{}):} \textit{``Refers to the problems found in software project documentation and can be identified by looking for missing, inadequate, or incomplete documentation of any type''} \cite{alves_towards_2014}. \documentationCount{} instances of \textit{documentation} debt were present. Several examples are presented below:
\begin{itemize}[itemsep=-2pt]
    \item \href{https://github.com/hibernate/hibernate-orm/blob/ff5fa10867d82d78002746faba4665e1a420bf0f/hibernate-core/src/test/java/org/hibernate/orm/test/annotations/cascade/circle/sequence/B.java\#L22}{\textcolor{commentColor}{/**
 * No Documentation
 */}}
 \item \href{https://github.com/ReactiveX/RxJava/blob/51cf3f4de0791fd1b0a7bd14bb6884d94c879612/src/test/java/io/reactivex/rxjava3/validators/ParamValidationNaming.java\#L324}{\textcolor{commentColor}{// no JavaDoc}}
 \item \href{https://github.com/elastic/elasticsearch/blob/3d457ef84ebf5d3ee0fd09339fbc3373ed5beb3c/server/src/test/java/org/elasticsearch/index/engine/InternalEngineTests.java\#L2175}{\textcolor{commentColor}{// TODO: add support for non-existing docs}}
 \item \href{https://github.com/hibernate/hibernate-orm/blob/ff5fa10867d82d78002746faba4665e1a420bf0f/hibernate-core/src/test/java/org/hibernate/orm/test/jpa/cascade2/ParentInfoAssigned.java\#L8}{\textcolor{commentColor}{/**
 * todo: describe ChildInfo
 *
 * @author Steve Ebersole
 */}}
 \item \href{https://github.com/hibernate/hibernate-orm/blob/ff5fa10867d82d78002746faba4665e1a420bf0f/hibernate-core/src/test/java/org/hibernate/orm/test/jpa/cascade2/Parent.java\#L8}{\textcolor{commentColor}{/**
 * todo: describe Parent
 *
 * @author Steve Ebersole
 */}}
\end{itemize}

\textbf{Design debt (\designCount{}):} \textit{``Refers to debt that can be discovered by analyzing the source code by identifying the use of practices which violated the principles of good object-oriented design (e.g. very large or tightly coupled classes)''} \cite{alves_towards_2014}. \designCount{} instances of design-related technical debt were identified, with representative examples shown below:
\begin{itemize}[itemsep=-2pt]
    \item \href{https://github.com/bazelbuild/bazel/blob/d7c71fa286b84a5dce8c60f1e8fb48295703fc61/src/test/java/com/google/devtools/build/lib/starlark/StarlarkRuleClassFunctionsTest.java\#L5441}{\textcolor{commentColor}{// TODO b/300201845 - encapsulate parents and childs private attributes}}
    \item \href{https://github.com/google/closure-compiler/blob/c86e4dd27ef48f9aa49261b1391274023ce7b658/test/com/google/javascript/jscomp/ExploitAssignsTest.java\#L223}{\textcolor{commentColor}{// TODO(b/123102446): We really like to collapse some of these chained assignments.}}
    \item \href{https://github.com/jetty/jetty.project/blob/12d16112fc65062f8d8de144a131b9f99dcae246/jetty-ee10/jetty-ee10-servlets/src/test/java/org/eclipse/jetty/ee10/servlets/GzipHandlerTest.java\#L42}{\textcolor{commentColor}{// TODO move to core}}
    \item \href{https://github.com/opensearch-project/OpenSearch/blob/c82fffe8e333a8fff600dc22fa2e1aed7276d180/server/src/test/java/org/opensearch/index/shard/SegmentReplicationWithNodeToNodeIndexShardTests.java\#L664}{\textcolor{commentColor}{// Todo: Move this test to SegmentReplicationIndexShardTests so that it runs for both node-node \& remote store}}
    \item \href{https://github.com/google/j2objc/blob/3664859a54dbe07905b2789cf06318c5b390e92f/jre_emul/android/platform/libcore/support/src/test/java/tests/support/Support_ASimpleWriter.java\#L6}{\textcolor{commentColor}{/**\\
 * An implementation of {@code OutputStream} that should serve as the\\
 * underlying stream for classes to be tested.\\
 * In particular this implementation allows to have IOExecptions thrown on demand.\\
 * For simplicity of use and understanding all fields are public.\\
 */}}
\end{itemize}

\textbf{Build debt (\buildCount{}):} \textit{``Refers to build related issues that make this task harder, and more time/processing consuming unnecessarily''} \cite{alves_towards_2014}. \buildCount{} instances of build-related technical debt were identified. Instances of build debt are as follows:

\begin{itemize}[itemsep=-2pt]
    \item \href{https://github.com/languagetool-org/languagetool/blob/a611c39ca0e3ee34ceb4f95bfc387aa9755566c8/languagetool-core/src/test/java/org/languagetool/tools/LtThreadPoolFactoryTest.java\#L227}{\textcolor{commentColor}{//Could fail if CI is to slow and will slow down the CI build; test local}}
    \item \href{https://github.com/google/j2objc/blob/3664859a54dbe07905b2789cf06318c5b390e92f/jre_emul/android/platform/libcore/harmony-tests/src/test/java/org/apache/harmony/tests/java/nio/channels/ServerSocketChannelTest.java\#L500}{\textcolor{commentColor}{// Continuous build environment doesn't support non-blocking IO.}}
    \item \href{https://github.com/geoserver/geoserver/blob/8871198a41d255de8c964263f0a6f15d44d15d62/src/extension/wps/wps-core/src/test/java/org/geoserver/wps/InputLimitsTest.java\#L611}{\textcolor{commentColor}{// see [GEOS-7835] WPS execution time limits test randomly breaks builds}}
    \item \href{https://github.com/jetty/jetty.project/blob/12d16112fc65062f8d8de144a131b9f99dcae246/jetty-ee9/jetty-ee9-demos/jetty-ee9-demo-async-rest/jetty-ee9-demo-async-rest-server/src/test/java/org/eclipse/jetty/ee9/demos/AsyncRestServer.java\#L28}{\textcolor{commentColor}{// Find the async-reset webapp based on common IDE working directories
// TODO import webapp as maven artifact}}
\end{itemize}

\textbf{Other debt (\otherCount{}):} This is the most generic form of SATD, and, in general, we labelled an SATD as other only if the SATD did not fit with any other specific type. A total of \otherCount{} instances of other debt were identified, with approximately 77\% of them beginning with the \textit{TODO} marker—a pattern that is not very common in other categories. Representative examples of other debt are shown below:
\begin{itemize}[itemsep=-2pt]
    \item \href{https://github.com/jetty/jetty.project/blob/12d16112fc65062f8d8de144a131b9f99dcae246/jetty-core/jetty-security/src/test/java/org/eclipse/jetty/security/SPNEGOAuthenticatorTest.java\#L98}{\textcolor{commentColor}{// TODO this test needs a lot of work}}
    \item \href{https://github.com/langchain4j/langchain4j/blob/0e5b1ece9cb22f52dfe9846885a35ec906e0cdd6/langchain4j-bedrock/src/test/java/dev/langchain4j/model/bedrock/common/BedrockChatModelIT.java\#L77}{\textcolor{commentColor}{// TODO implement}}
    \item \href{https://github.com/elastic/elasticsearch/blob/3d457ef84ebf5d3ee0fd09339fbc3373ed5beb3c/x-pack/plugin/core/src/test/java/org/elasticsearch/xpack/core/ml/action/GetDatafeedsActionRequestTests.java\#L25}{\textcolor{commentColor}{// TODO implement https://github.com/elastic/\newline elasticsearch/issues/25929}}
    \item \href{https://github.com/NationalSecurityAgency/ghidra/blob/db1b6860c9bbe8b659704cb3909c0bf54359fb7d/Ghidra/Features/Base/src/test.slow/java/ghidra/app/util/bin/format/dwarf/DWARFDataTypeImporterTest.java\#L1197}{\textcolor{commentColor}{// not implemented yet}}
\end{itemize}



\begin{tcolorbox}
\textbf{\underline{RQ1 summary}:} 
We identified \categoryCount{} SATD types in test code, including \newCategoryCount{} new test-specific categories such as skipped and limited tests. Also, for the categories that were found in source code SATD, the distributions are significantly different than ours. These findings reveal that test code exhibits unique technical debt patterns not fully captured by existing classifications from application code.
\end{tcolorbox}

\subsection{RQ2: Can existing SATD detection tools detect test code SATD?}
\textbf{Motivation.} As prior research has shown that self-admitted technical debt (SATD) adversely affects software quality and maintenance efforts~\cite{chowdhury_evidence_2025}, the ability to automatically detect SATD is essential for taking timely actions. Yet, it is unknown whether current automated techniques can identify SATD within test code, which we investigate here.

\textbf{Approach.} We employed seven well-established SATD detection approaches originally developed for source code comments, following the same implementations used in prior comparative work whenever possible.

\textbf{Pattern.} We use the 62 keyword/phrase patterns summarized by Potdar et al.~\cite{potdar_exploratory_2014}. For each target comment, if any predefined pattern is matched, the comment is labeled as SATD; otherwise, it is labeled as non-SATD. This is a strict rule-based unsupervised method and does not require model training.

\textbf{MAT.} Similar to Pattern, Matches task Annotation Tags (MAT)~\cite{guo_how_2021} is unsupervised; however, it differs in two key ways: it is centered on widely used task-annotation tags (e.g., TODO/FIXME-style tags supported by IDEs), and it applies fuzzy matching rather than strict exact matching, enabling more flexible identification of SATD-like expressions in comments.

\textbf{SATD Detector.} We include the text-mining based SATD Detector by Liu et al.~\cite{liu_satd_2018}, a supervised classifier that identifies SATD comments from textual features through preprocessing, feature selection, and vector-space representations.

\textbf{TM.} We also include another text-mining approach by Huang et al.~\cite{huang_identifying_2018}, for which we use the implementation packaged by Guo et al.~\cite{guo_how_2021}.

\textbf{NLP.} Maldonado et al.~\cite{maldonado_using_2017} formulated SATD detection as an NLP-based text classification task with a Maximum Entropy classifier. As in Guo et al.~\cite{guo_how_2021}, we rely on the available reimplementation (the original code was not released) to ensure reproducibility and a consistent experimental pipeline.

\textbf{MT-BERT.} For the multitask BERT model proposed by Gu et al.~\cite{gu_self-admitted_2024}, we use the original released implementation with default settings reported by the authors. BERT is a bidirectional transformer-based language model that learns contextual representations from text.

\textbf{DebtHunter.} We also evaluate DebtHunter~\cite{sala_debthunter_2021}, a machine-learning-based SATD detection approach that classifies comments using learned textual features.

In summary, Pattern, SATD Detector, NLP, TM, and MAT are executed via Guo et al.'s JAR package~\cite{guo_how_2021}, while MT-BERT and DebtHunter are run using their original implementations~\cite{gu_self-admitted_2024,sala_debthunter_2021}. For supervised methods (SATD Detector, NLP, TM, MT-BERT, and DebtHunter), we report pretrained, retrained, and retrained-with-cross-validation settings where applicable. We report Precision, Recall, F1-score, and Matthews Correlation Coefficient (MCC).

\begin{table*}[ht!]
\centering
\caption{SATD detection performance with/without retraining. The unsupervised MAT approach achieved the highest F1-score (also, NLP~\cite{maldonado_using_2017} but only with cross-validation). Pattern-based methods had high precision but very low recall, and the Text Mining (TM) Method showed low precision but relatively higher recall. We also show the minimum, maximum, and average of all the scores across all the tools and approaches.}
\label{table:non_llm_satd_detection_result}
\begin{tabular}{cccccccc}
\toprule
\textbf{Dataset} & \textbf{Method} & \textbf{Retrained?} & \textbf{Cross-Validation?} & \textbf{Precision} & \textbf{Recall} & \textbf{F1-score} & \textbf{MCC}\\

\midrule
\multirow{17}{*}{\rotatebox[origin=c]{45}{Original}}
  & Pattern~\cite{potdar_exploratory_2014} & -- & --    &   0.80 & 0.03 & 0.06 & 0.15\\
  & MAT~\cite{guo_how_2021} & -- & --    &   \textbf{0.95} & 0.61 & \textbf{0.75} & \textbf{0.76}\\
  & SATD Detector~\cite{liu_satd_2018} & \xmark & \xmark    &   0.38 & 0.58 & 0.46 & 0.46\\
  & SATD Detector~\cite{liu_satd_2018} & \checkmark & \xmark    &   0.30 & 0.57 & 0.40 & 0.41\\
  & SATD Detector~\cite{liu_satd_2018} & \checkmark & \checkmark    &   0.35 & 0.63 & 0.44 & 0.45\\
  & TM~\cite{huang_identifying_2018} & \xmark & \xmark    &   0.09 & 0.56 & 0.15 & 0.20\\
  & TM~\cite{huang_identifying_2018} & \checkmark & \xmark    &   0.10 & 0.86 & 0.17 & 0.26\\
  & TM~\cite{huang_identifying_2018} & \checkmark & \checkmark    &   0.10 & \textbf{0.88} & 0.18 & 0.28\\
  & NLP~\cite{maldonado_using_2017} & \xmark & \xmark    &   0.53 & 0.54 & 0.53 & 0.53\\
  & NLP~\cite{maldonado_using_2017} & \checkmark & \xmark    &   0.68 & 0.63 & 0.65 & 0.65\\
  & NLP~\cite{maldonado_using_2017} & \checkmark & \checkmark    &   0.86 & 0.67 & \textbf{0.75} & 0.75\\
  & MT-BERT~\cite{gu_self-admitted_2024} & \xmark & \xmark    &   0.83 & 0.63 & 0.71 & 0.72\\
  & MT-BERT~\cite{gu_self-admitted_2024} & \checkmark & \xmark    &   0.63 & 0.50 & 0.56 & 0.56\\
  & MT-BERT~\cite{gu_self-admitted_2024} & \checkmark & \checkmark    &   0.68 & 0.45 & 0.54 & 0.55\\
  & DebtHunter~\cite{sala_debthunter_2021} & \xmark & \xmark    &   0.26 & 0.58 & 0.36 & 0.38\\
  & DebtHunter~\cite{sala_debthunter_2021} & \checkmark & \xmark    &  0.53 & 0.65 & 0.59 & 0.58\\
  & DebtHunter~\cite{sala_debthunter_2021} & \checkmark & \checkmark    & 0.68 & 0.69 & 0.68 & 0.68\\
  & \textit{Minimum} & -- & -- & 0.09 & 0.03 & 0.06 & 0.15\\
  & \textit{Maximum} & -- & -- & 0.95 & 0.88 & 0.75 & 0.76\\
  & \textit{Average} & -- & -- & 0.52 & 0.59 & 0.47 & 0.49\\
\midrule
\multirow{17}{*}{\rotatebox[origin=c]{45}{Deduplicate}}
  & Pattern~\cite{potdar_exploratory_2014} & -- & --    &   0.80 & 0.04 & 0.07 & 0.17\\
  & MAT~\cite{guo_how_2021} & -- & --    &   \textbf{0.95} & 0.56 & \textbf{0.70} & \textbf{0.73}\\
  & SATD Detector~\cite{liu_satd_2018} & \xmark & \xmark    &   0.37 & 0.63 & 0.46 & 0.47\\
  & SATD Detector~\cite{liu_satd_2018} & \checkmark & \xmark    &   0.31 & 0.61 & 0.41 & 0.42\\
  & SATD Detector~\cite{liu_satd_2018} & \checkmark & \checkmark    &   0.36 & 0.65 & 0.46 & 0.47\\
  & TM~\cite{huang_identifying_2018} & \xmark & \xmark    &   0.14 & 0.49 & 0.22 & 0.24\\
  & TM~\cite{huang_identifying_2018} & \checkmark & \xmark    &   0.28 & 0.62 & 0.39 & 0.40\\
  & TM~\cite{huang_identifying_2018} & \checkmark & \checkmark    &   0.30 & 0.63 & 0.40 & 0.42\\
  & NLP~\cite{maldonado_using_2017} & \xmark & \xmark    &   0.49 & 0.50 & 0.49 & 0.48\\
  & NLP~\cite{maldonado_using_2017} & \checkmark & \xmark    &   0.64 & 0.58 & 0.61 & 0.60\\
  & NLP~\cite{maldonado_using_2017} & \checkmark & \checkmark    &   0.81 & 0.62 & \textbf{0.70} & 0.70\\
  & MT-BERT~\cite{gu_self-admitted_2024} & \xmark & \xmark    &   0.83 & 0.58 & 0.69 & 0.69\\
  & MT-BERT~\cite{gu_self-admitted_2024} & \checkmark & \xmark    &   0.65 & 0.43 & 0.52 & 0.53\\
  & MT-BERT~\cite{gu_self-admitted_2024} & \checkmark & \checkmark    &   0.60 & 0.41 & 0.49 & 0.49\\
  & DebtHunter~\cite{sala_debthunter_2021} & \xmark & \xmark    &  0.34 & 0.53 & 0.41 & 0.41\\
  & DebtHunter~\cite{sala_debthunter_2021} & \checkmark & \xmark    &  0.21 & 0.65 & 0.32 & 0.35\\
  & DebtHunter~\cite{sala_debthunter_2021} & \checkmark & \checkmark    &  0.47 & \textbf{0.69} & 0.54 & 0.55\\
  & \textit{Minimum} & -- & -- & 0.14 & 0.04 & 0.07 & 0.17\\
  & \textit{Maximum} & -- & -- & 0.95 & 0.69 & 0.70 & 0.73\\
  & \textit{Average} & -- & -- & 0.50 & 0.54 & 0.46 & 0.48\\
\bottomrule
\end{tabular}

\end{table*}

\textbf{Results.} The detection results of the seven approaches are summarized in Table~\ref{table:non_llm_satd_detection_result}. The pattern-based approach consistently achieved high precision (0.80) but extremely low recall on both the original (0.03) and deduplicated (0.04) test sets, resulting in very low F1-scores (0.06 and 0.07, respectively). Notably, both precision and F1-score are approximately two to four times lower for our test code dataset than for their original dataset. This outcome highlights the approach’s limited coverage for test code, as it relies on a fixed set of 62 manually curated patterns obtained through a source code-dominated dataset. The low recall can be attributed, in part, to missing keywords in the manually crafted list, specifically the absence of \texttt{TODO} alongside \texttt{FIX}, as well as the lack of discernible patterns in many comments. 

The MAT approach achieved the best overall performance across both datasets without any training. It consistently yielded high precision (0.95) and achieved the highest F1-scores on both the original (0.75) and deduplicated (0.70) datasets, with corresponding recall values of 0.61 and 0.56. Notably, both MAT and the NLP approach achieved the highest F1-scores, with NLP attaining this performance under the cross-validation setting. Nevertheless, compared to prior work, the F1-score observed in our evaluation is 2\% lower.

The two text-mining approaches exhibit notably different performance profiles on test code comments. Compared with Huang et al.’s TM approach, SATD Detector consistently achieves much higher precision, F1-score, and MCC on both datasets, while maintaining competitive recall. However, SATD Detector remains clearly below the retrained NLP approach, which attains substantially stronger balanced performance, indicating that SATD Detector provides better coverage-oriented performance than TM but not the same overall effectiveness as NLP. Meanwhile, the TM approach produced consistently low precision and F1-scores across all settings. Similar to the NLP pretrained model, TM failed to recognize \texttt{TODO}, although it correctly identified \texttt{FIXME}, and produced numerous false positives from patterns such as issue IDs (e.g., \texttt{// DROOLS-155}), copyright notices, and Javadocs with author annotations. After retraining, TM corrected many of these patterns but still misclassified a substantial number of irrelevant comments as SATD. Nevertheless, its recall was comparable to that of the NLP and MAT approaches, and was even the highest for the retrained cross-validation setting on the original dataset. Compared to Huang et al.’s~\cite{huang_identifying_2018} reported F1-score of 0.737 for source code, its performance on test code is notably poor.

For both datasets, the retrained NLP model achieved roughly 12\% higher F1-scores than the pretrained model, producing balanced precision and recall. On the pretrained model, many comments containing the two major keywords, \texttt{TODO} and \texttt{FIXME}, were missed, and the misidentified comments generally lacked clear patterns. After retraining, the model successfully learned \texttt{TODO} but still failed to capture \texttt{FIXME}. Notably, most of the remaining misclassified comments were Javadocs describing bug-related tests that include bug ID (e.g., \texttt{@bug 8077931}). Compared to Guo et al.’s~\cite{guo_how_2021} result on source code, the best F1-score in our case after retraining is about 5\% lower.

The MT-BERT approach exhibited strong performance on pretrained, achieving F1-scores of 0.71 and 0.69 on the original and deduplicated test datasets, respectively. This indicates that the pretrained multitask model generalizes reasonably well to test code comments. However, retraining MT-BERT on the test code comment datasets resulted in a noticeable performance degradation, primarily due to reduced recall, which led to lower F1-scores. Manual inspection revealed that MT-BERT attained relatively high precision by correctly identifying SATD comments containing explicit debt-related keywords; nevertheless, it occasionally failed to detect when common keywords such as \texttt{TODO} were present, and similar to MAT, it struggled with comments lacking explicit keyword signals, leading to comparable recall between the two approaches. Overall, the pretrained version performed better in these cases, likely because it was trained on a larger and more diverse set of SATD patterns, enabling it to better capture keyword-based and recurring SATD expressions. Although DebtHunter improved significantly after retraining, its accuracy on test code is much lower than its reported accuracy on source code~\cite{sala_debthunter_2021}, and its lower precision kept its overall performance below the best MAT and NLP results.

The overall poor to moderate performance indicates that methods originally developed based on the characteristics of source code fail to fully capture the distinctive properties of test code. This finding reinforces the notion that SATD instances in test code differ substantially from those in source code. Although retraining sometimes improved the performance, even in the best scenarios, results remained 2–5\% lower than those on the source code. These findings highlight the somewhat poor transferability of source-trained models and show that retraining alone is not sufficient.

\begin{tcolorbox}
\textbf{\underline{RQ2 summary}:} 
Although the NLP and MAT approaches achieve the highest accuracy in detecting SATD in test code, their F1-scores remain relatively modest at around 0.75. This limitation motivated us to investigate the effectiveness of different LLMs for detecting SATD in test code, particularly because prior research has demonstrated the strong performance of LLMs across various software engineering tasks, including the detection of SATD in source code.
\end{tcolorbox}
\subsection{RQ3: Can open-source LLMs detect SATD in test code?}
\textbf{Motivation.} Although SATD often contains keywords such as \textit{TODO}, many comments do not follow such patterns. In these cases, even sophisticated pattern-based methods (e.g., TM, NLP) fail to detect SATD, highlighting the need for approaches that capture semantic meaning. Consequently, we investigate Large Language Models (LLMs), given their strong capability in understanding both natural language and source code~\cite{brown_language_2020}. Recent studies have also found promising results with LLMs for detecting source code SATD~\cite{sheikhaei_empirical_2024, lambert_identification_2024}, outperforming conventional approaches.
\begin{figure}
    \centering
    \includegraphics[width=\columnwidth]{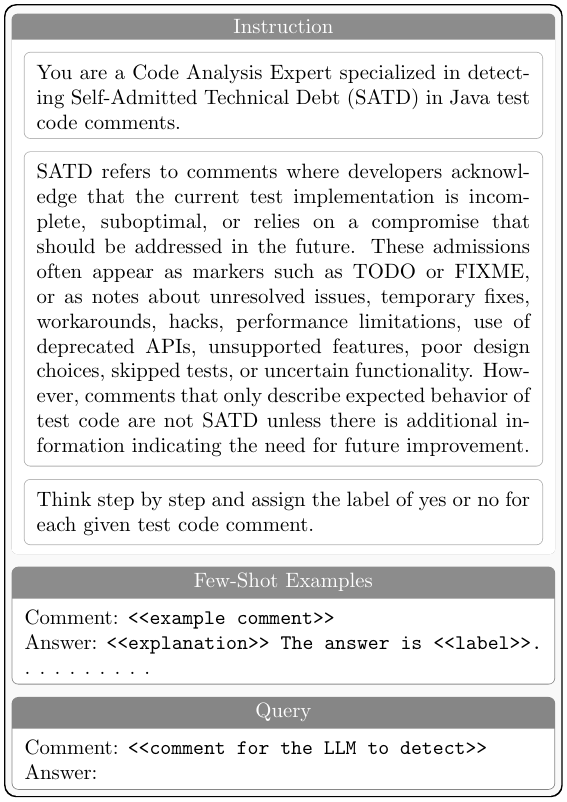}
    \caption{LLM prompt template for detecting SATD comment.}
    \label{fig:satd-detection-prompt-template}
\end{figure}

\textbf{Approach.} Encouraged by the success of SATD detection~\cite{sheikhaei_empirical_2024} of the Flan-T5 model series, we investigated these models on test code SATD with few-shot configurations of 0, 2, and 4 shots. These models are T5-SMALL, T5-BASE, T5-LARGE, T5-XL, and T5-XXL, with numbers of parameters 60M to 11B. The experiments were run on an HPC cluster with Intel 8570 CPUs (2.1 GHz) and multiple NVIDIA H100 GPUs. We tested four different prompt templates: No Keyword, MAT, Jitterbug, and GPT-4. These were adopted from Sheikhaei et al.~\cite{sheikhaei_empirical_2024} to enable direct comparison with their results. 

Additionally, we added a new prompt template that we mainly developed for RQ4, as illustrated in Figure~\ref{fig:satd-detection-prompt-template}. Existing research highlights that LLMs are highly sensitive to the structure and phrasing of prompts, with their performance varying considerably depending on how tasks, instructions, and examples are articulated~\cite{liu_pre-train_2023}. To ensure greater reliability and adaptability, a modular prompt template was developed, comprising three components, in accordance with established prompt design frameworks~\cite{amatriain_prompt_2024, giray_prompt_2023}. The first, the Instruction, is organized into three parts: (i) a role-based persona that anchors the model’s behavior and enhances response quality~\cite{hu_quantifying_2024}; (ii) a contextual and task-specific description that clarifies the model’s objective~\cite{amatriain_prompt_2024}; and (iii) explicit output formatting guidelines that reduce ambiguity and improve consistency in responses. The second, Few-Shot Examples, varies in number across experiments to assess performance sensitivity. Finally, the third, Query, contains the target comment that the LLM is required to analyze and classify.

To assess the stability of the Flan-T5 results, we independently repeated each experimental configuration three times under identical conditions. Each repetition used the same test data, prompt, few-shot examples, model, and inference settings. We used deterministic decoding rather than probabilistic sampling; therefore, parameters that control the randomness and diversity of generated responses were not applied. Since Flan-T5 occasionally generated explanatory text in addition to the requested answer, we extracted the final prediction label, such as \texttt{yes} or \texttt{no}, from each response and treated it as the prediction. We then compared the outputs across the three runs.
\begin{table*}[ht!]
\centering
\caption{ Precision (P), Recall (R), F1-score (F1), and Matthews Correlation Coefficient (MCC) for detecting SATD comments using Flan-T5 series open-source LLMs across 0-, 2-, and 4-shot settings. Keyword-based prompts (e.g., MAT, Jitterbug) generally achieve higher F1-scores, whereas smaller models and No Keyword/GPT-4–suggested prompts often fail to identify SATD.}
\label{table:oss_llm_detection_result}

\begin{tabular}{p{0.8cm}cccccccccccccc}
\toprule
\multirow{2}{*}{\textbf{Dataset}} & \multirow{2}{*}{\textbf{Prompt}} & \multirow{2}{*}{\textbf{Model}} 
& \multicolumn{4}{c}{\textbf{0-shot}} 
& \multicolumn{4}{c}{\textbf{2-shots}} 
& \multicolumn{4}{c}{\textbf{4-shots}} \\
\cmidrule(lr){4-7} \cmidrule(lr){8-11} \cmidrule(lr){12-15}
 &  &  & \textbf{P} & \textbf{R} & \textbf{F1} & \textbf{MCC}
 & \textbf{P} & \textbf{R} & \textbf{F1} & \textbf{MCC} 
 & \textbf{P} & \textbf{R} & \textbf{F1} & \textbf{MCC} \\ 
\midrule
\multirow{25}{*}{\rotatebox[origin=c]{90}{Original}}&\multirow{5}{*}{\centering Ours} & T5-SMALL  & 0.00 & 0.02 & 0.01 & -0.02  & 0.00 & 0.00 & 0.00 & -0.01  & 0.00 & 0.00 & 0.00 & -0.01  \\
& & T5-BASE  & 0.01 & 0.31 & 0.02 & -0.03  & 0.02 & 0.51 & 0.03 & 0.01  & 0.01 & 0.45 & 0.03 & 0.01  \\
& & T5-LARGE  & 0.03 & 0.96 & 0.05 & 0.10  & 0.04 & 0.91 & 0.07 & 0.14  & 0.04 & 0.74 & 0.07 & 0.12  \\
& & T5-XL  & 0.29 & 0.86 & 0.43 & 0.49  & 0.38 & 0.79 & 0.51 & 0.54  & 0.47 & 0.64 & 0.54 & 0.54  \\
& & T5-XXL  & 0.10 & \textbf{0.98} & 0.18 & 0.29  & 0.12 & 0.96 & 0.21 & 0.32  & 0.11 & 0.95 & 0.20 & 0.31  \\
\cmidrule(lr){2-15}
&\multirow{5}{*}{\centering \makecell{No\\Keyword}} & T5-SMALL  & 0.00 & 0.00 & 0.00 & 0.00  & 0.00 & 0.00 & 0.00 & 0.00  & 0.00 & 0.00 & 0.00 & 0.00  \\
& & T5-BASE  & 0.00 & 0.03 & 0.01 & -0.04  & 0.00 & 0.00 & 0.00 & 0.00  & 0.00 & 0.00 & 0.00 & -0.00  \\
& & T5-LARGE  & 0.00 & 0.00 & 0.00 & 0.00  & 0.00 & 0.00 & 0.00 & 0.00  & 0.00 & 0.00 & 0.00 & 0.00  \\
& & T5-XL  & 0.33 & 0.01 & 0.01 & 0.05  & 0.00 & 0.00 & 0.00 & 0.00  & 0.00 & 0.00 & 0.00 & 0.00  \\
& & T5-XXL  & 0.10 & 0.66 & 0.18 & 0.24  & 0.19 & 0.50 & 0.28 & 0.29  & 0.17 & 0.50 & 0.25 & 0.28  \\
\cmidrule(lr){2-15}
&\multirow{5}{*}{\centering GPT 4} & T5-SMALL  & 0.00 & 0.00 & 0.00 & 0.00  & 0.00 & 0.00 & 0.00 & 0.00  & 0.00 & 0.00 & 0.00 & 0.00  \\
& & T5-BASE  & 0.01 & 0.04 & 0.01 & -0.03  & 0.00 & 0.00 & 0.00 & -0.00  & 0.00 & 0.00 & 0.00 & -0.02  \\
& & T5-LARGE  & 0.24 & 0.04 & 0.06 & 0.09  & 0.29 & 0.01 & 0.03 & 0.06  & 0.67 & 0.01 & 0.03 & 0.10  \\
& & T5-XL  & \textbf{1.00} & 0.04 & 0.07 & 0.19  & 0.00 & 0.00 & 0.00 & 0.00  & 0.00 & 0.00 & 0.00 & 0.00  \\
& & T5-XXL  & 0.20 & 0.87 & 0.32 & 0.40  & 0.18 & 0.82 & 0.30 & 0.37  & 0.18 & 0.80 & 0.30 & 0.36  \\
\cmidrule(lr){2-15}
&\multirow{5}{*}{\centering Jitterbug} & T5-SMALL  & 0.00 & 0.00 & 0.00 & 0.00  & 0.00 & 0.00 & 0.00 & 0.00  & 0.00 & 0.00 & 0.00 & 0.00  \\
& & T5-BASE  & 0.00 & 0.03 & 0.01 & -0.03  & 0.00 & 0.00 & 0.00 & -0.00  & 0.00 & 0.00 & 0.00 & -0.01  \\
& & T5-LARGE  & 0.00 & 0.00 & 0.00 & 0.00  & 0.00 & 0.00 & 0.00 & 0.00  & 0.00 & 0.00 & 0.00 & 0.00  \\
& & T5-XL  & \textbf{1.00} & 0.47 & 0.64 & 0.68  & \textbf{1.00} & 0.11 & 0.20 & 0.33  & \textbf{1.00} & 0.08 & 0.15 & 0.28  \\
& & T5-XXL  & 0.28 & 0.83 & 0.42 & 0.47  & 0.27 & 0.82 & 0.41 & 0.46  & 0.30 & 0.80 & 0.43 & 0.48  \\
\cmidrule(lr){2-15}
&\multirow{5}{*}{\centering MAT} & T5-SMALL  & 0.00 & 0.00 & 0.00 & 0.00  & 0.00 & 0.00 & 0.00 & 0.00  & 0.00 & 0.00 & 0.00 & 0.00  \\
& & T5-BASE  & 0.00 & 0.04 & 0.01 & -0.03  & 0.00 & 0.00 & 0.00 & -0.00  & 0.00 & 0.00 & 0.00 & -0.00  \\
& & T5-LARGE  & 0.00 & 0.00 & 0.00 & 0.00  & 0.00 & 0.00 & 0.00 & 0.00  & 0.00 & 0.00 & 0.00 & 0.00  \\
& & T5-XL  & 0.99 & 0.52 & \textbf{0.68} & \textbf{0.71}  & \textbf{1.00} & 0.13 & 0.23 & 0.36  & \textbf{1.00} & 0.12 & 0.21 & 0.34  \\
& & T5-XXL  & 0.33 & 0.80 & 0.47 & 0.50  & 0.40 & 0.78 & 0.53 & 0.55  & 0.42 & 0.74 & 0.54 & 0.55  \\
\midrule
\multirow{25}{*}{\rotatebox[origin=c]{90}{Deduplicate}}&\multirow{5}{*}{\centering Ours} & T5-SMALL  & 0.01 & 0.03 & 0.01 & -0.02  & 0.00 & 0.00 & 0.00 & -0.01  & 0.00 & 0.00 & 0.00 & -0.01  \\
& & T5-BASE  & 0.02 & 0.36 & 0.04 & 0.01  & 0.02 & 0.55 & 0.04 & 0.03  & 0.02 & 0.52 & 0.04 & 0.03  \\
& & T5-LARGE  & 0.03 & 0.95 & 0.05 & 0.09  & 0.04 & 0.89 & 0.08 & 0.14  & 0.05 & 0.76 & 0.09 & 0.14  \\
& & T5-XL  & 0.26 & 0.85 & 0.40 & 0.46  & 0.36 & 0.76 & 0.49 & 0.51  & 0.46 & 0.65 & 0.54 & 0.53  \\
& & T5-XXL  & 0.10 & \textbf{0.97} & 0.18 & 0.28  & 0.11 & \textbf{0.97} & 0.20 & 0.31  & 0.11 & 0.96 & 0.20 & 0.30  \\
\cmidrule(lr){2-15}
&\multirow{5}{*}{\centering \makecell{No\\Keyword}} & T5-SMALL  & 0.00 & 0.00 & 0.00 & 0.00  & 0.00 & 0.00 & 0.00 & 0.00  & 0.00 & 0.00 & 0.00 & 0.00  \\
& & T5-BASE  & 0.01 & 0.04 & 0.02 & -0.01  & 0.00 & 0.00 & 0.00 & 0.00  & 0.00 & 0.00 & 0.00 & -0.00  \\
& & T5-LARGE  & 0.00 & 0.00 & 0.00 & 0.00  & 0.00 & 0.00 & 0.00 & 0.00  & 0.00 & 0.00 & 0.00 & 0.00  \\
& & T5-XL  & 0.00 & 0.00 & 0.00 & -0.00  & 0.00 & 0.00 & 0.00 & 0.00  & 0.00 & 0.00 & 0.00 & 0.00  \\
& & T5-XXL  & 0.11 & 0.74 & 0.20 & 0.26  & 0.22 & 0.58 & 0.32 & 0.34  & 0.21 & 0.57 & 0.31 & 0.33  \\
\cmidrule(lr){2-15}
&\multirow{5}{*}{\centering GPT 4} & T5-SMALL  & 0.00 & 0.00 & 0.00 & 0.00  & 0.00 & 0.00 & 0.00 & 0.00  & 0.00 & 0.00 & 0.00 & 0.00  \\
& & T5-BASE  & 0.01 & 0.04 & 0.02 & -0.01  & 0.00 & 0.00 & 0.00 & -0.00  & 0.00 & 0.00 & 0.00 & -0.01  \\
& & T5-LARGE  & 0.25 & 0.04 & 0.07 & 0.10  & 0.29 & 0.02 & 0.03 & 0.07  & 0.67 & 0.02 & 0.03 & 0.11  \\
& & T5-XL  & \textbf{1.00} & 0.04 & 0.07 & 0.19  & 0.00 & 0.00 & 0.00 & 0.00  & 0.00 & 0.00 & 0.00 & 0.00  \\
& & T5-XXL  & 0.19 & 0.87 & 0.31 & 0.39  & 0.18 & 0.87 & 0.29 & 0.37  & 0.18 & 0.83 & 0.30 & 0.37  \\
\cmidrule(lr){2-15}
&\multirow{5}{*}{\centering Jitterbug} & T5-SMALL  & 0.00 & 0.00 & 0.00 & 0.00  & 0.00 & 0.00 & 0.00 & 0.00  & 0.00 & 0.00 & 0.00 & 0.00  \\
& & T5-BASE  & 0.01 & 0.02 & 0.01 & -0.01  & 0.00 & 0.00 & 0.00 & -0.00  & 0.00 & 0.00 & 0.00 & -0.01  \\
& & T5-LARGE  & 0.00 & 0.00 & 0.00 & 0.00  & 0.00 & 0.00 & 0.00 & 0.00  & 0.00 & 0.00 & 0.00 & 0.00  \\
& & T5-XL  & \textbf{1.00} & 0.40 & 0.57 & 0.63  & \textbf{1.00} & 0.12 & 0.22 & 0.35  & \textbf{1.00} & 0.09 & 0.16 & 0.29  \\
& & T5-XXL  & 0.28 & 0.82 & 0.42 & 0.47  & 0.26 & 0.81 & 0.39 & 0.44  & 0.28 & 0.79 & 0.41 & 0.45  \\
\cmidrule(lr){2-15}
&\multirow{5}{*}{\centering MAT} & T5-SMALL  & 0.00 & 0.00 & 0.00 & 0.00  & 0.00 & 0.00 & 0.00 & 0.00  & 0.00 & 0.00 & 0.00 & 0.00  \\
& & T5-BASE  & 0.01 & 0.03 & 0.01 & -0.01  & 0.00 & 0.00 & 0.00 & -0.00  & 0.00 & 0.00 & 0.00 & -0.00  \\
& & T5-LARGE  & 0.00 & 0.00 & 0.00 & 0.00  & 0.00 & 0.00 & 0.00 & 0.00  & 0.00 & 0.00 & 0.00 & 0.00  \\
& & T5-XL  & \textbf{1.00} & 0.46 & \textbf{0.63} & \textbf{0.68}  & \textbf{1.00} & 0.15 & 0.26 & 0.39  & \textbf{1.00} & 0.13 & 0.23 & 0.36  \\
& & T5-XXL  & 0.34 & 0.79 & 0.47 & 0.50  & 0.40 & 0.76 & 0.52 & 0.54  & 0.40 & 0.72 & 0.52 & 0.53  \\
\bottomrule

\end{tabular}%
\end{table*}

\textbf{Results.} All three runs produced the same classification results for every experimental configuration. We therefore report the results from the first run. Table~\ref{table:oss_llm_detection_result} reports the results for the T5 model family, ranging from T5-SMALL to T5-XXL. The findings indicate that the Flan-T5 series LLMs are unable to consistently detect SATD in test code, with their effectiveness varying depending on the prompting strategy, shots, and model size. In most cases, the smaller models---T5-Small, T5-Base, and T5-Large---failed to detect SATD, resulting in zero recall and, consequently, zero precision.

Between the two largest models, T5-XL consistently achieved higher precision than T5-XXL, often reaching near-perfect precision (e.g., 1.0) with keyword-based prompts (MAT and Jitterbug). However, this high precision generally came at the expense of very low recall. The Ours prompt performed competitively, achieving a maximum F1-score of 0.54. In contrast, the No Keyword and GPT-4 prompts yielded poor results, with many zero-shot and few-shot runs failing to detect any SATD (F1 = 0.0). Meanwhile, the T5-XXL models demonstrated higher recall, frequently exceeding 0.8, but with substantially lower precision (ranging from 0.10 to 0.42). 

When comparing the F1-scores of T5-XXL across different prompt variations with those reported by Sheikhaei et al.~\cite{sheikhaei_empirical_2024}, we observe similar overall trends; however, the scores in our evaluation are approximately 30\% lower. Overall, the MAT-based keyword prompts achieved the highest F1-score (0.68). This strong performance is largely due to the high prevalence of MAT-suggested keywords—particularly \textit{TODO}—in SATD comments and their absence in non-SATD comments within our test dataset. For instance, using the MAT approach, the model correctly identified 71 SATD comments out of 137 with the Flan-T5-XL (zero-shot) setup but failed to detect the remaining 66, even when other obvious indicators (e.g., yuck, workaround) were present. A clear example of a missed case is the comment \textit{“// kinda sorta wrong, but for testing’s sake...”}, which expresses technical debt but lacks any MAT-suggested keywords.

These findings indicate that the choice of prompt has a substantial impact on model performance. Increasing the number of shots (0→2→4), however, did not consistently improve results; this trend was also observed by Sheikhaei et al.~\cite{sheikhaei_empirical_2024}. In both studies, the best performance was achieved under zero-shot settings, as demonstrated by T5-XL with MAT-based prompts in our experiments. Project-level analysis confirmed that T5-XL with the MAT prompt in the zero-shot setting significantly outperformed the other configurations ($p < 0.05$) in both F1-score and MCC, with non-negligible effect sizes.

\begin{tcolorbox}
\textbf{\underline{RQ3 summary}:} 
The results indicate that open-source LLMs (Flan-T5 series) struggle to reliably detect SATD in test code across different prompt types and shot settings. Although the best F1-score reached 0.68 using the MAT prompt, the overall performance remains inconsistent and relatively poor compared to the existing tools (RQ2), suggesting that these models have a limited ability to understand and generalize SATD patterns in test code.
\end{tcolorbox}

\subsection{RQ4: Can proprietary LLMs detect SATD in test code?}
\textbf{Motivation.} Motivated by the unexpectedly poor performance of open-source LLMs observed in RQ3, we extended our investigation to proprietary LLMs. Previous research consistently reports that proprietary models tend to outperform open-source counterparts across a range of standard benchmarks, including those that involve code understanding and generation tasks~\cite{dvivedi2024comparative, mayer_can_2024, jimenez2023swe}.

\textbf{Approach.} We evaluated two major proprietary LLM families—Gemini and GPT—to assess their effectiveness in detecting SATD in test code. For Gemini, we used the latest Flash variants of Gemini 2.0 and 2.5, and for GPT, we tested three GPT-5 models (nano, mini, and standard) to capture diverse reasoning capabilities and model sizes. Each model was tested under the few-shot configurations of 0, 2, and 4 shots. All the experiments were conducted using the prompt shown in Figure~\ref{fig:satd-detection-prompt-template}. 

\begin{table*}[!h]
\centering
\caption{Precision (P), Recall (R), F1-score (F1) and Matthews Correlation Coefficient (MCC) for detecting SATD comments using GPT and Gemini across 0-, 2-, and 4-shot settings.}
\label{table:proprietary_llm_detection_result}
\begin{tabular}{cccccccccccccc}
\toprule
\multirow{2}{*}{\textbf{Dataset}} & \multirow{2}{*}{\textbf{Model}} 
& \multicolumn{4}{c}{\textbf{0-shot}} 
& \multicolumn{4}{c}{\textbf{2-shots}} 
& \multicolumn{4}{c}{\textbf{4-shots}} \\
\cmidrule{3-14}
 &  & \textbf{P} & \textbf{R} & \textbf{F1} & \textbf{MCC} 
 & \textbf{P} & \textbf{R} & \textbf{F1} & \textbf{MCC} 
 & \textbf{P} & \textbf{R} & \textbf{F1} & \textbf{MCC}  \\
\midrule

\multirow{5}{*}{\rotatebox[origin=c]{45}{Original}} & gpt-5-nano  & 0.12 & 0.93 & 0.20 & 0.31  & 0.18 & 0.91 & 0.30 & 0.39  & 0.19 & 0.91 & 0.31 & 0.40  \\
 & gpt-5-mini  & 0.23 & 0.95 & 0.36 & 0.45  & 0.25 & 0.93 & 0.39 & 0.47  & 0.26 & 0.96 & 0.40 & 0.48  \\
 & gpt-5  & 0.27 & 0.95 & 0.42 & 0.50  & \textbf{0.34} & 0.93 & \textbf{0.49} & \textbf{0.55}  & 0.32 & 0.94 & 0.47 & 0.54  \\
\cmidrule(lr){2-2} \cmidrule(lr){3-14}
 & gemini-2.5-flash  & 0.15 & \textbf{0.99} & 0.26 & 0.36  & 0.17 & \textbf{1.00} & 0.28 & 0.39  & 0.12 & \textbf{0.99} & 0.21 & 0.33  \\
 & gemini-2.0-flash  & 0.24 & 0.88 & 0.37 & 0.44  & 0.15 & 0.98 & 0.26 & 0.37  & 0.12 & \textbf{0.99} & 0.21 & 0.32  \\
\midrule
\multirow{5}{*}{\rotatebox[origin=c]{45}{Deduplicate}} & gpt-5-nano  & 0.11 & 0.92 & 0.20 & 0.30  & 0.19 & 0.91 & 0.31 & 0.39  & 0.17 & 0.89 & 0.29 & 0.38  \\
 & gpt-5-mini  & 0.22 & 0.94 & 0.35 & 0.44  & 0.24 & 0.94 & 0.39 & 0.47  & 0.24 & 0.96 & 0.38 & 0.46  \\
 & gpt-5  & 0.27 & 0.96 & 0.42 & 0.49  & \textbf{0.31} & 0.92 & \textbf{0.46} & \textbf{0.52}  & 0.30 & 0.93 & 0.45 & \textbf{0.51}  \\
\cmidrule(lr){2-2} \cmidrule(lr){3-14}
 & gemini-2.5-flash  & 0.15 & \textbf{0.99} & 0.26 & 0.36  & 0.17 & \textbf{1.00} & 0.29 & 0.39  & 0.12 & \textbf{0.99} & 0.21 & 0.31  \\
 & gemini-2.0-flash  & 0.24 & 0.88 & 0.38 & 0.45  & 0.15 & 0.98 & 0.27 & 0.37  & 0.11 & \textbf{0.99} & 0.20 & 0.31  \\

\bottomrule

\end{tabular}
\end{table*}

\textbf{Results.} The experimental results are summarized in Table~\ref{table:proprietary_llm_detection_result}. Proprietary models achieved consistently high recall, frequently exceeding 0.90 and occasionally reaching 1.0 (e.g., Gemini-2.5-Flash under 2-shot). However, this high recall was accompanied by very low precision, particularly for Gemini models, where precision often ranged between 0.11 and 0.24. Among GPT models, the standard GPT-5 demonstrated the most balanced performance, attaining the highest F1-scores under 2-shot settings on both the original (0.49) and deduplicated (0.46) datasets. The GPT-5 mini and nano variants performed moderately well, outperforming Gemini in terms of precision and F1, but lagging behind GPT-5 standard. Increasing the number of shots did not consistently improve performance, as was seen with the Flan-T5 models as well. Similar to the open-source LLMs, the project-level analysis showed that the best-performing configuration, GPT-5 in the 2-shot setting, significantly outperformed the other configurations ($p < 0.05$) in both F1-score and MCC, with non-negligible effect sizes.

Compared to the Flan-T5 series (RQ3), proprietary models exhibit distinct performance patterns. GPT-5 models maintain high recall, unlike T5-XL, which achieves strong precision primarily with MAT or Jitterbug prompts but suffers from limited recall. GPT-5 provides more stable SATD detection across shot settings compared to the Flan-T5 series. Nonetheless, Flan-T5-XL with MAT prompts achieved the highest F1-score (0.68).


\begin{tcolorbox}
\textbf{\underline{RQ4 summary}:}
In general, proprietary LLMs, particularly GPT-5, outperform open-source models in recall. However, their much lower precision indicates a tendency to over-predict SATD instances. These results suggest that while proprietary models may offer stronger contextual reasoning to exhibit high recall, significant improvements are needed to address their limitations with precision.
\end{tcolorbox}

\section{Discussion}\label{sec:discussion}
SATD can significantly hinder software maintenance activities. Understanding the nature of SATD and its detection is, therefore, critical for effective mitigation. This study aimed to investigate the landscape of SATD in test code—an area largely overlooked by previous research, despite the critical role of testing in software development and the well-documented impact of SATD on software maintainability. In this paper, we examined the types of SATD that appear in test code (RQ1) and found \categoryCount{} types, where \newCategoryCount{} of them were new---not found in other source code-based studies. We then investigated the effectiveness of the existing tools (RQ2), open-source LLMs (RQ3), and proprietary LLMs (RQ4) in identifying these SATD comments. Our evaluation revealed that existing tools, particularly MAT, are able to detect SATD with high precision (0.95), but their recall ($\leq 0.61$) remains limited. Meanwhile, open-source LLMs exhibited inconsistent and, in some cases, unexpectedly poor performance across model sizes and prompt variations. In the best cases, such as Flan-T5-XL with the MAT prompt, high precision but low recall was observed. In contrast, proprietary LLMs exhibited very high recall but very low precision.

While we leave a complete systematic qualitative analysis of the inaccurate detection of test code SATD by all studied tools and approaches as future work, we still sought to gain an initial understanding of the root causes of these failures. To this end, the first author, who has over 10 years of industry experience, manually analyzed the best-performing configuration in each of the three approach families: MAT among existing tools, Flan-T5-XL with the MAT prompt in the zero-shot setting among open-source LLMs, and GPT-5 in the 2-shot setting among proprietary LLMs. For the best-performing existing approach, MAT, the first author manually analyzed all 50 missed SATD comments in the deduplicated test set. They found that, although many of these false negatives do express technical debt, they do so through test-specific wording rather than the source-code SATD cues that existing approaches were largely designed around. Representative examples include the following:

\begin{itemize}[itemsep=-2pt]
    \item \href{https://github.com/trinodb/trino/blob/b89da7e651fea169443b1aaedb0c406744fe2cff/core/trino-main/src/test/java/io/trino/sql/query/TestInlineFunctions.java\#L423}{\textcolor{commentColor}{// inline function cannot be recursive
// note: mutual recursion is not supported either, but it is not tested due to the forward declaration limitation above}}
    \item \href{https://github.com/apache/druid/blob/58f31b821fe9802ba73f7bbc3234c629ad73f299/sql/src/test/java/org/apache/druid/sql/calcite/CalciteExportTest.java\#L229}{\textcolor{commentColor}{// Disabled until replace supports external destinations. To be enabled after that point.}}
    \item \href{https://github.com/trinodb/trino/blob/b89da7e651fea169443b1aaedb0c406744fe2cff/plugin/trino-base-jdbc/src/test/java/io/trino/plugin/jdbc/BaseJdbcTableStatisticsTest.java\#L293}{\textcolor{commentColor}{// Not testing average length and min/max, as this would make the test less reusable and is not that important to test.}}
\end{itemize}

These examples indicate that lower recall is concentrated in test-specific wordings or expressions that are common in test SATD comments. Even when comments contain words such as \texttt{disabled}, \texttt{deactivated}, or \texttt{skip}, these expressions function as test-specific cues rather than canonical source-code SATD markers such as \texttt{TODO} or \texttt{FIXME}. As a result, keyword- and pattern-dependent approaches can preserve high precision while still missing a large portion of genuine SATD comments.

For open-source LLMs, the same author manually analyzed all 61 SATD comments missed by Flan-T5-XL in the MAT-based zero-shot setting. Again, many missed comments relied on test-specific phrasing rather than conventional SATD keywords. Although these models are more flexible than strict rule-based techniques, their strongest results still come from keyword-oriented prompts such as MAT and Jitterbug, suggesting that they remain strongly anchored to lexical cues inherited from source-code SATD. Examples include:

\begin{itemize}[itemsep=-2pt]
    \item\href{https://github.com/openjdk/jdk/blob/642816538fbaa5b74c6beb8a14d1738cdde28c10/test/langtools/tools/javac/lambdaShapes/org/openjdk/tests/vm/FDSeparateCompilationTest.java\#L24}{\textcolor{commentColor}{// this test has been disabled because of timeout issues.
// see JDK-8006746}}
    \item \href{https://github.com/bazelbuild/bazel/blob/d7c71fa286b84a5dce8c60f1e8fb48295703fc61/src/test/java/com/google/devtools/build/lib/bazel/repository/downloader/HttpConnectorTest.java\#L281}{\textcolor{commentColor}{// Deactivated due to https://github.com/bazelbuild\\/bazel/issues/9380.}}
    \item \href{https://github.com/JodaOrg/joda-time/blob/15c3df573457a34e36d4b918e114d9c2d7e45396/src/test/java/org/joda/time/format/TestDateTimeFormatStyle.java\#L159}{\textcolor{commentColor}{// skip test, as it needs historical time zone info}}
\end{itemize}

To our surprise, the proprietary LLMs exhibited the opposite behavior: very high recall but very low precision. To understand their lower precision, we randomly sampled 146 of the 233 non-SATD comments misclassified as SATD (95\% confidence level) by GPT-5 in the 2-shot setting and manually analyzed their characteristics. In these cases, the model often overinterpreted generic task-oriented, cautionary, or context-setting comments as technical debt. Representative false positives include:

\begin{itemize}[itemsep=-2pt]
    \item \href{https://github.com/flowable/flowable-engine/blob/bb50957ef795e979a9fb31f18de39a5788fa5a9b/modules/flowable-engine/src/test/java/org/flowable/engine/test/bpmn/gateway/InclusiveGatewayTest.java\#L1231}{\textcolor{commentColor}{//Finish pending tasks}}
    \item \href{https://github.com/helidon-io/helidon/blob/cb510cf63346c42ec2206c2b4f02727a3e0f7243/integrations/cdi/jpa-cdi/src/test/java/io/helidon/integrations/cdi/jpa/TestAlternatePersistenceXmlLocation.java\#L53}{\textcolor{commentColor}{// Load it dynamically because Hibernate won't be on the classpath when we're testing with Eclipselink}}
    \item \href{https://github.com/apache/druid/blob/58f31b821fe9802ba73f7bbc3234c629ad73f299/integration-tests-ex/cases/src/test/java/org/apache/druid/testsEx/auth/ITSecurityBasicQuery.java\#L309}{\textcolor{commentColor}{// Wait for a second so that the auth is synced, to avoid flakiness}}
\end{itemize}

These examples help explain why proprietary LLMs achieve high recall but low precision. Such models appear to use broader semantic reasoning, which allows them to recover SATD comments that keyword-dependent approaches miss. However, the same behavior can cause them to overinterpret ordinary explanatory comments as debt. In the first example, the phrase \texttt{pending tasks} can be read as unresolved work, even though the comment merely describes the purpose of the following statement. In the second example, the model may interpret the dependency-related wording as evidence of an architectural limitation or workaround, although the comment only explains a test setup decision under a specific classpath condition. In the third example, the phrasing resembles a temporary workaround for unstable behavior, which can sound like admitted debt, but the comment is still describing test synchronization logic rather than explicitly admitting postponed maintenance. Taken together, these cases show that broader semantic reasoning improves recall, but it can also weaken discrimination between SATD and non-SATD comments when operational, dependency-related, or test-stabilization comments resemble debt-related language. More generally, the quantitative and qualitative results indicate that SATD in test code differs from source-code SATD not only in category but also in how they are generally expressed. Reliable detection of SATD in test comments will therefore likely require models that learn test-specific cues in context, rather than relying only on source-code SATD keywords or on overly broad semantic interpretations.

Future work can explore additional open-source models, incorporate richer contextual information (e.g., surrounding code), and train or fine-tune models to improve detection accuracy. In particular, future training data and prompts should explicitly capture test-specific SATD expressions and keywords such as \texttt{disabled}, \texttt{deactivated}, and \texttt{skip}, rather than relying mainly on source-code-oriented cues such as \texttt{TODO} and \texttt{FIXME}. Including such test-specific cues, together with hard negative examples that use similar operational language without actually admitting debt, may improve both recall and precision. When such accurate detection is possible, we can investigate the evolution of test methods with SATD~\cite{grund_codeshovel_2021, chowdhury_evidence_2025} to understand their long-term maintenance impact (e.g., change- and bug-proneness of SATD test methods compared to non-SATD test methods). We hope that our manually curated dataset of test code SATD will encourage many future research projects focusing on automated test code SATD detection, classification, and the impact of test code SATD in long-term software maintenance.


\section{Threats to Validity}\label{sec:threats}

\textbf{Construct validity.} Our SATD detection and classification were completely based on manual analysis. Since manual interpretation is inherently susceptible to personal bias and subjectivity, we involved multiple reviewers to mitigate this risk.

\textbf{Internal validity.} Internal validity is hampered by our selection of LLMs. However, the open-source LLMs were selected based on their excellent performance in a similar task. While selecting the proprietary LLMs, we focused on the prominent Gemini and GPT models that exhibited good performance in many software engineering tasks~\cite{batista2024code, bruni2025benchmarking, simoes2024evaluating}. LLM outputs may vary across executions. Although we mitigated this threat by repeating the experiments with open-source LLMs, we did not repeat the experiments with proprietary LLMs due to the associated computational costs.

\textbf{External validity.} To enhance the generalizability of our findings, we analyzed 1,000 open-source projects spanning a variety of domains. However, all selected projects were written in Java. As a result, the findings may not generalize to closed-source projects or software developed in other programming languages. Also, despite analyzing a large number of comments (50K), the relatively low number of SATD comments in the manually reviewed dataset remains a limitation; expanding this dataset could enhance the robustness of our findings.

\textbf{Conclusion validity} of this study is hampered by all the aforementioned threats. 
\section{Conclusion}\label{sec:conclusion}
This study presents a large-scale empirical investigation of Self-Admitted Technical Debt (SATD) in test code. Through manual analysis of 50,000 comments from 1,000 open-source Java projects, we identified \totalSatdCount{} SATD instances, which were further categorized into \categoryCount{} types, including four newly proposed ones. The findings reveal that SATD in test code exhibits distinct characteristics compared to SATD in source code. Our evaluation of SATD detection approaches further demonstrates that existing tools achieve high precision but suffer from low recall, open-source LLMs display inconsistent performance across models and prompt configurations, and proprietary LLMs attain higher recall at the expense of precision. We anticipate that these insights will inspire future research and facilitate the development of more robust, and accurate techniques for the automated detection and classification of SATD. Ultimately, such advancements have the potential to support practical adoption in industry, enabling developers to better manage technical debt and reduce long-term maintenance overhead.

\section*{Acknowledgment}
This research was supported by an NSERC Discovery Grant and a University of Manitoba start-up grant.


\bibliographystyle{elsarticle-num}
\bibliography{references, references-more}

\end{document}